\documentclass[fleqn,usenatbib]{mnras}

\usepackage{newtxtext,newtxmath}

\usepackage[T1]{fontenc}
\usepackage{pdflscape}

\DeclareRobustCommand{\VAN}[3]{#2}
\let\VANthebibliography\thebibliography
\def\thebibliography{\DeclareRobustCommand{\VAN}[3]{##3}\VANthebibliography}

\usepackage{graphicx}	
\usepackage{amsmath}	
\usepackage{booktabs,tabularx,makecell}

\newcommand{\oii}{O\,{\scriptsize II}}
\newcommand{\oiii}{O\,{\scriptsize III}}

\newcommand{\ciii}{C\,{\scriptsize III}}

\newcommand{\niii}{N\,{\scriptsize III}}
\newcommand{\niv}{N\,{\scriptsize IV}}
\newcommand{\neiii}{Ne\,{\scriptsize III}}
\newcommand{\nv}{N\,{\scriptsize V}}
\newcommand{\civ}{C\,{\scriptsize IV}}

\newcommand{\heii}{He\,{\scriptsize II}}

\newcommand{\lya}{Ly$\alpha$}

\usepackage{rotating}
\usepackage{float}
\usepackage{amsmath}
\usepackage{graphicx} 
\usepackage[caption=false]{subfig}
\usepackage{makecell}

\graphicspath{{./}{figures/}}

\title[The Spectroscopic Diversity of Galaxies in the First 500 Myr]{JWST Spectroscopic Insights Into the Diversity of Galaxies in the First 500 Myr: Short-Lived Snapshots Along a Common Evolutionary Pathway}

\author[G. Roberts-Borsani et al.]
{Guido Roberts-Borsani,$^{1}$\thanks{E-mail: g.robertsborsani@ucl.ac.uk} 
Pascal A. Oesch,$^{2,3,4}$
Richard Ellis,$^{1}$
Andrea Weibel,$^{2}$
Emma Giovinazzo,$^{2}$
\newauthor
Rychard Bouwens,$^{5}$
Pratika Dayal,$^{6,7}$
Adriano Fontana,$^{8}$
Kasper E. Heintz,$^{3,9,2}$
Jorryt Matthee,$^{10}$
\newauthor
Romain A. Meyer,$^{2}$
Laura Pentericci,$^{8}$
Alice Shapley,$^{11}$
Sandro Tacchella,$^{12,13}$
Tommaso Treu,$^{11}$
\newauthor
Fabian Walter,$^{14,15}$
Hakim Atek,$^{16}$
Sownak Bose,$^{17}$
Marco Castellano,$^{8}$
Yoshinobu Fudamoto,$^{18}$
\newauthor
Takahiro Morishita,$^{19,20}$
Rohan P. Naidu,$^{21}$
Ryan L. Sanders,$^{22}$
and
Arjen van der Wel$^{23}$
\\
$^{1}$Department of Physics \& Astronomy, University College London, London, WC1E 6BT, UK \\
$^{2}$Department of Astronomy, University of Geneva, Chemin Pegasi 51, 1290 Versoix, Switzerland \\
$^{3}$Cosmic Dawn Center (DAWN), Denmark \\
$^{4}$Niels Bohr Institute, University of Copenhagen, Jagtvej 128, København N, DK-2200, Denmark \\
$^{5}$Leiden Observatory, Einsteinweg 55, NL-2333 CC Leiden, The Netherlands \\
$^{6}$Kapteyn Astronomical Institute, University of Groningen, PO Box 800, 9700 AV Groningen, The Netherlands \\
$^{7}$Canadian Institute for Theoretical Astrophysics, 60 St George St, University of Toronto, Toronto, ON M5S 3H8, Canada \\
$^{8}$INAF - Osservatorio Astronomico di Roma, Via Frascati 33, 00078 Monteporzio Catone, Rome, Italy \\
$^{9}$Niels Bohr Institute, University of Copenhagen, Jagtvej 128, 2200 Copenhagen N, Denmark \\
$^{10}$Institute of Science and Technology Austria (ISTA), Am Campus 1, 3400 Klosterneuburg, Austria \\
$^{11}$Department of Physics \& Astronomy, University of California, Los Angeles, 430 Portola Plaza, Los Angeles, CA 90095, USA \\
$^{12}$Kavli Institute for Cosmology, University of Cambridge, Madingley Road, Cambridge, CB3 0HA, UK \\
$^{13}$Cavendish Laboratory, University of Cambridge, 19 JJ Thomson Avenue, Cambridge, CB3 0HE, UK \\
$^{14}$Max Planck Institut f\"ur Astronomie, K\"onigstuhl 17, D-69117 Heidelberg, Germany \\
$^{15}$California Institute of Technology, Pasadena, CA 91125, USA \\
$^{16}$Institut d’Astrophysique de Paris, UMR 7095, CNRS, and Sorbonne Universit\'e, 98 bis boulevard Arago, 75014 Paris, France \\
$^{17}$Institute for Computational Cosmology, Department of Physics, Durham University, South Road, Durham DH1 3LE, UK \\
$^{18}$Center for Frontier Science, Chiba University, 1-33 Yayoi-cho, Inage-ku, Chiba 263-8522, Japan \\
$^{19}$IPAC, California Institute of Technology, MC 314-6, 1200 E. California Boulevard, Pasadena, CA 91125, USA \\
$^{20}$Astronomical Institute, Tohoku University, 6-3 Aramaki, Aoba-ku, Sendai 980-8578, Japan \\
$^{21}$MIT Kavli Institute for Astrophysics and Space Research, 70 Vassar Street, Cambridge, MA 02139, USA \\
$^{22}$Department of Physics and Astronomy, University of Kentucky, 505 Rose Street, Lexington, KY 40506, USA \\
$^{23}$Sterrenkundig Observatorium, Universiteit Gent, Krijgslaan 281 S9, 9000 Gent, Belgium
}

\date{Accepted XXX. Received YYY; in original form ZZZ}

\pubyear{\the\year{}}
\graphicspath{{./}{figures/}}
\begin{document}
\label{firstpage}
\pagerange{\pageref{firstpage}--\pageref{lastpage}}
\maketitle

\begin{abstract}
We investigate the nature and spectroscopic diversity of early galaxies from a sample of 41 sources at $z\geqslant10$ with JWST/NIRSpec prism observations. We compare the properties of strong UV line emitters, traced by intense \civ\ emission, with those of more ``typical'' sources with weak or undetected \civ. The more typical (or ``\civ-weak'') sources reveal significant scatter in their \ciii] line strengths, UV continuum slopes, and physical sizes, spanning \ciii] equivalent widths of $\sim$1-51 \AA, UV slopes of $\beta\sim-1.6$ to $-2.6$, and half-light radii of $\sim$50-1000 pc. In contrast, \civ-strong sources occupy the tail of these distributions, with \ciii] EWs of 16-51 \AA, UV slopes $\beta\lesssim-2.5$, compact morphologies ($r_{\rm 50} \lesssim 100$\,pc), and elevated star formation surface densities ($\Sigma_{\rm SFR} \gtrsim 100\,M_\odot\,\mathrm{yr}^{-1}\,\mathrm{kpc}^{-2}$). These properties suggest concentrated starbursts that temporarily outshine the host galaxy. Comparing average properties from composite spectra, we find the diversity of the sample is primarily driven by bursty star formation on very short timescales ($\leq$3 Myr), with strong \civ\ emitters observed at the apex of the bursts and sources devoid of emission lines during relative inactivity. An apparent association between strong \civ\ and enhanced nitrogen abundance suggests both may be modulated by the same duty cycle, reflecting a generic mode of star formation. We show that AGN are unlikely to contribute significantly to this duty cycle based on UV line diagnostics and photoionisation models. Our results support a picture whereby brief bursts and lulls can explain the spectral diversity and early growth of bright galaxies in the first 500 Myr.
\end{abstract}




\section{Introduction}
The study of galaxies out to the earliest of times affords a unique window into the initial conditions that set the stage for the formation and evolution of the chemical abundances, supermassive black holes (SMBHs), and large-scale structures we see today \citep{dayal18,issi24,stark25}. Accordingly, direct measurements of the earliest sources represents the most viable approach towards constraining galaxy formation and chemical enrichment models. Until recently, the necessary data to do so has been sparse, owing to the limited sensitivity and wavelength coverage of ground-based spectroscopy and Hubble Space Telescope (HST) imaging, restricting such studies to sources below redshifts $z\sim10-11$ (e.g., 
\citealt{bouwens11,coe13,dunlop13,ellis13,labbe13,smit14,bouwens15,finkelstein15,oesch16,rb16,laporte17,laporte17b,stark17,bowler20,rb20,rb22,tacchella22}).

The impressive capabilities of the James Webb Space Telescope (JWST) have now confidently extended this frontier, enabling photometric measurements out to $z\sim15-20$ (e.g., \citealt{donnan24,robertson24,castellano25,perezgonzalez25,whitler25,weibel25}) and, crucially, spectroscopic measurements to comparable redshifts reaching $\sim300$ Myr after the Big Bang (e.g., \citealt{bunker23,castellano24,napolitano25,carniani24,curtislake23,witstok25,naidu25}).  Such early measurements proved significant, revealing a number of extraordinary spectra characterised by apparently anomalous properties not generally seen in typical lower-redshift sources (c.f. \citealt{castellano24,bunker23,napolitano24,topping25,rb24,hayes25,deugenio23,scholtz25,kokorev25}). In particular, the presence of strong carbon (\ciii]$\lambda\lambda$1907,1909 \AA\ and \civ$\lambda\lambda$1548,1550 \AA) and/or nitrogen emission (\niii]$\lambda\lambda$1747,1749 \AA\ and \niv]$\lambda$1486 \AA) in a handful of $z\geqslant10$ sources revealed a class of galaxies characterised by extremely hard ionising fields, metal-poor interstellar media (ISM), and enhanced nitrogen-to-oxygen (N/O) abundances possibly tracing (super)massive stars and/or a variable initial mass function \citep{charbonnel23,senchyna24,rui24,topping24}.

Although the primary physical mechanisms driving such remarkable features remain debated, it is also clear that many of the $z\geqslant10$ spectra studied thus far do not show such extreme line emission, highlighting a contrast with more typical sources and an important yet less explored diversity among the galaxy population in the first $\simeq$500 Myr \citep{arrabal23,harikane25,naidu25,castellano24,napolitano24}.
As such, a logical question arises: Are these unusual sources a distinct population tracing atypical physical processes and evolutionary pathways, or do they represent brief snapshots of extreme activity within a common evolutionary pathway? Furthermore, what astrophysical processes regulate the spread of observed $z\geqslant10$ galaxy properties to give rise to such diversity?

Answering these questions serves as a fundamental test of our understanding of the initial phases of galaxy evolution and requires deep spectroscopy over enlarged $z\geqslant10$ samples in order to determine: (i) the prevalence of anomalous conditions as traced by particularly high ionisation lines such as \civ\ and \niv] line emission, (ii) the physics governing the underlying radiation fields, stellar populations, and ISM of the general population, and (iii) the timescales on which they act within the context of the star formation history (SFH) associated to the typical $z\geqslant10$ source. The remarkably efficient capabilities of NIRSpec have now made the confirmation and study of $z\sim10-14$ sources practical (e.g., \citealt{tang25}), affording the larger samples necessary to answer these questions and gain contextual insight into the conditions powering the average spectrum of galaxies within the first 500 Myr after the Big Bang.

The primary aim of this paper is to achieve these overarching goals. To do so, we leverage the largest available sample of $z\geqslant10$ spectra to date, including those of newly-confirmed sources, to investigate the spectral diversity of $z\geqslant10$ galaxies traced by individual sources and average composite spectra, placing strong \civ\ and \niv] line emitters into context relative to the more typical luminous population and extending observational templates out to $z\simeq10-14$. Specifically, this paper aims to characterise the observable features accessible to NIRSpec and NIRCam for such high redshift galaxies (e.g., emission line strengths, UV continuum slopes, half-light radii), in order to constrain their ISM and stellar conditions (e.g., metallicities, dust contents, stellar ages, recent star formation activity, and assembly histories). The paper is structured as follows. Section~\ref{sec:data} presents the JWST data sets utilised in this study, the compilation of the largest $z\geqslant10$ spectroscopic sample to date, and the procedures by which we derive composite spectra and measurements of spectroscopic quantities. In Section~\ref{sec:distribution} we present these measurements and observables for the full sample of $z\geqslant10$ galaxies, including a characterisation of their distributions and comparison between strong and weak UV line emitters. Section~\ref{sec:props} presents the distributions of inferred galaxy properties from the constraints presented in Section~\ref{sec:distribution}, and Section~\ref{sec:diversity_disc} sets these results into context with a discussion on the main driver(s) behind the spectroscopic diversity of $z\geqslant10$ galaxies. A summary of our findings and conclusions is presented in Section~\ref{sec:summary}. Throughout this paper we adopt a cosmology with $H_{0}=67\,\rm{km}\,\rm{s}^{-1}\,\rm{Mpc}^{-1}$, $\Omega_{\rm M}=0.3$, $\Omega_{\Lambda}=0.7$ \citep{planck20}. All magnitudes are quoted in the AB system \citep{oke83}.

\section{Data sets and Methods}
\label{sec:data}
\subsection{NIRSpec Prism Spectroscopy and NIRCam Photometry}
\label{subsec:reduction}
We utilise JWST NIRSpec prism spectroscopy (nominally covering $\approx$0.6-5.3 $\mu$m with spectral resolution $R\sim30-300$) from virtually all publicly available Multi-Shutter Assembly (MSA) observations, as well as a small number of proprietary data sets covering a mix of blank fields and lensing clusters. The data reduction process and 1D spectral extractions are conducted with version 0.9.8 of the \texttt{msaexp} code \citep{msaexp} and follow the procedures broadly outlined by \citet{rb24} and \citet{meyer25}, which adopt a local sky background and are effectively identical to the processes described by the RUBIES collaboration in \citet{degraaff24} (c.f. also with \citealt{heintz24}). We refer the interested reader to these papers for details. In total, the data reduction and spectral extraction processes result in $\sim24,500$ spectra for consideration. For sources where multiple observations exist (as indicated by multiple MSA IDs in Table~\ref{tab:confirmations}), we combine the spectra via an inverse-variance weighted mean for a deeper spectrum.

We also make use of NIRCam imaging for additional slit loss verifications and corrections (if and where required) as well as morphological analyses. For these, we utilise the photometric catalogs of \citet{weibel25}, which are based on 0.04 arcsec-per-pixel mosaics from the Dawn JWST Archive\footnote{\url{https://dawn-cph.github.io/dja/}} (DJA). For full details on the data reduction and photometric catalog we refer the reader to \citet{weibel25} and \citet{valentino23}. In fields not covered by the catalog, we supplement with photometry from the Astrodeep catalog \citep{merlin24}, or from the literature.

\subsection{$z\geqslant10$ Spectroscopic Sample}
Each of the spectra from our data reduction procedure
is fit with the EAzY \citep{brammer08} redshift-fitting component module of \texttt{msaexp} and visually inspected. A significant number of $z\geqslant10$ confirmations have already been reported in the literature (specifically, \citealt{arrabal23,arrabal23b,bunker23,curtislake23,harikane23,carniani24,castellano24,fujimoto24a,hainline24,hsiao24,kokorev25,naidu25,napolitano25,schouws25,tang25,weibel25,witstok25}) and we thus automatically include these a priori in our fiducial sample.
In addition to these sources, we present new spectroscopic redshifts for eight galaxies owing to the detection of a convincing Lyman-$\alpha$ break and/or emission lines (e.g., \civ$\lambda\lambda$1548,1550 \AA, \ciii]$\lambda\lambda$1907,1909 \AA, [\oii]$\lambda\lambda$3727,3729 \AA), and list these along with the rest of the sample in Table~\ref{tab:confirmations} -- three of these newly-confirmed sources derive from observations over the UDS and COSMOS fields as part of the ``Mirage of Miracle'' program (GO 5224, PIs Oesch \& Naidu) and will be presented in a forthcoming paper (Oesch et al, in prep.), while the remaining five come from archival observations over the EGS and GOODS-South fields as part of the CAPERS (GO 6368, PI Dickinson), CEERS (DDT 2750, PI Arrabal Haro), and JADES (GTO 1286 and 1287, PI Luetzgendorf and Isaak, respectively) programs, respectively. The full suite of program IDs upon which this sample is based, along with their associated survey papers, are listed in the Acknowledgements.

Given a non-negligible number of confirmations come from the Lyman-$\alpha$ break, which in principle can be confused with a lower-redshift Balmer break, we compare in Appendix~\ref{sec:photoz} the derived spectroscopic redshifts to the inferred photometric redshifts using \texttt{EAzY}, finding excellent agreement between the two in supporting a $z\geqslant10$ solution. Moreover, we also compare the spectroscopic redshifts to those inferred in previous works, also finding good agreement.

\begin{table*}
\centering
\caption{The confirmed $z\geqslant10$ sources utilised in this study, with basic spectro-photometric properties. All F200W fluxes are quoted without correction for gravitational lensing, where applicable. The initial spectroscopic confirmation of each source is referenced in the right column, with the following abbreviations: Naidu25=\citet{naidu25}, Car24=\citet{carniani24}, Sch24=\citet{schouws25}, Wit25b=\citet{witstok25}, CL23=\citet{curtislake23}. Cast24=\citet{castellano24}, Nap25=\citet{napolitano25}, Hsiao24=\citet{hsiao24}, AH23a=\citet{arrabal23}, AH23b=\citet{arrabal23b}, Fuj24=\citet{fujimoto24a}, Bun23=\citet{bunker23}, Sch25=\citet{schouws25}, Koko25=\citet{kokorev25}, Weibel25=\citet{weibel25}, Tang25=\citet{tang25}, Wit25a=\citet{witstok25b}.}
\label{tab:confirmations}
\begin{small}
\begin{tabularx}{\textwidth}{l X r r c X c c c X}
\toprule
Name & MSA-ID & RA & Dec & $z_{\rm spec}$ & Field & $\mu$ & F200W & $M_{\rm UV}$ & Ref. \\
\midrule
MoM-z14 & 5224\_277193 & 150.09333 & 2.27316 & 14.440 & COSMOS & -- & 28.51$\pm$0.25 & $-$20.00$\pm$0.11 & Naidu25 \\
GS-z14-0 & 1287\_183348 & 53.08294 & $-$27.85563 & 14.179 & GOODS-S & -- & 27.04$\pm$0.05 & $-$21.30$\pm$0.05 & Car24,Sch25 \\
GS-z14-1 & 1287\_20018044 & 53.07427 & $-$27.88592 & 14.044 & GOODS-S & -- & 28.87$\pm$0.23 & $-$19.21$\pm$0.15 & Car24 \\
JADES-GS-z13-1-LA & 1287\_20013731 & 53.06475 & $-$27.89024 & 12.955 & GOODS-S & -- & 29.20$\pm$0.19 & $-$18.70$\pm$0.19 & Wit25b \\
GS-z13-0 & 3215\_20128771 & 53.14988 & $-$27.77650 & 12.926 & GOODS-S & -- & 29.08$\pm$0.15 & $-$18.82$\pm$0.15 & CL23 \\
UNCOVER-13077 & 2561\_13077 & 3.57087 & $-$30.40159 & 12.908 & Abell 2744 & 2.5 & 27.53$\pm$0.11 & $-$19.37$\pm$0.11 & Fuj24 \\
GS-z12-0 & 3215\_20096216 & 53.16635 & $-$27.82156 & 12.516 & GOODS-S & -- & 28.78$\pm$0.11 & $-$19.07$\pm$0.11 & CL23 \\
UNCOVER-38766 & 2561\_38766 & 3.51356 & $-$30.35680 & 12.393 & Abell 2744 & 1.7 & 28.16$\pm$0.17 & $-$19.10$\pm$0.17 & Fuj24 \\
CAPERS-EGS-65480 & 6368\_65480 & 214.80717 & 52.79020 & 12.344 & EGS & -- & 28.36$\pm$0.26 & $-$19.48$\pm$0.26 & This Work \\
GHZ2/GLASS-z12 & 3073\_22600 & 3.49898 & $-$30.32475 & 12.338 & Abell 2744 & 1.3 & 26.59$\pm$0.05 & $-$20.96$\pm$0.05 & Cast24 \\
CEERS-1 & 2750\_1 & 214.94315 & 52.94244 & 11.400 & EGS & -- & 27.47$\pm$0.11 & $-$20.25$\pm$0.11 & AH23a \\
JADES-GS-20015720 & 1287\_20015720 & 53.11763 & $-$27.88818 & 11.267 & GOODS-S & -- & 28.14$\pm$0.17 & $-$19.56$\pm$0.17 & Tang25 \\
GS-z11-0 & \makecell[l]{1210\_14220,\\3215\_20130158} & 53.16477 & $-$27.77463 & 11.122 & GOODS-S & -- & 28.29$\pm$0.11 & $-$19.38$\pm$0.11 & CL23,Wit25a \\
CEERS-10 & 2750\_10 & 214.90663 & 52.94550 & 11.040 & EGS & -- & 27.41$\pm$0.12 & $-$20.26$\pm$0.12 & AH23a \\
CAPERS-UDS-z11 & 6368\_126973 & 34.26444 & $-$5.09623 & 11.010 & UDS & -- & 26.77$\pm$0.15 & $-$20.93$\pm$0.15 & Koko25 \\
MoM-z11-1 & 5224\_10132 & 150.11405 & 2.19134 & 10.921 & COSMOS & -- & 27.68$\pm$0.24 & $-$19.98$\pm$0.24 & This Work \\
JADES-GS-20177294 & 1287\_20177294 & 53.07900 & $-$27.86359 & 10.893 & GOODS-S & -- & 28.34$\pm$0.09 & $-$19.25$\pm$0.09 & Tang25 \\
CAPERS-EGS-43539 & 6368\_43539 & 214.90310 & 52.80659 & 10.816 & EGS & -- & 28.82$\pm$0.36 & $-$18.82$\pm$0.36 & This Work \\
MoM-z11-2 & 5224\_154407 & 34.39431 & $-$5.12917 & 10.803 & UDS & -- & 27.62$\pm$0.22 & $-$20.02$\pm$0.22 & This Work \\
EGS-22637 & \makecell[l]{1345\_80073,\\6368\_22637} & 214.93206 & 52.84187 & 10.704 & EGS & -- & 27.40$\pm$0.12 & $-$20.23$\pm$0.12 & This Work \\
GHZ4 & 3073\_16115 & 3.51374 & $-$30.35157 & 10.698 & Abell 2744 & 1.6 & 27.64$\pm$0.14 & $-$19.47$\pm$0.14 & Nap25 \\
JADES-GS-20176151 & \makecell[l]{1287\_20176151,\\1286\_20176151} & 53.07076 & $-$27.86544 & 10.619 & GOODS-S & -- & 28.11$\pm$0.14 & $-$19.50$\pm$0.14 & Tang25 \\
EGS-69 & 2750\_69 & 214.86160 & 52.90460 & 10.619 & EGS & -- & 27.83$\pm$0.21 & $-$19.78$\pm$0.21 & This Work \\
GN-z11 & \makecell[l]{1181\_3991,\\1211\_1268} & 189.10605 & 62.24205 & 10.611 & GOODS-N & -- & 26.07$\pm$0.01 & $-$21.54$\pm$0.01 & Bun23 \\
CAPERS-UDS-z10 & 6368\_136645 & 34.45602 & $-$5.12195 & 10.581 & UDS & -- & 27.53$\pm$0.18 & $-$20.08$\pm$0.18 & Koko25 \\
GHZ7 & 3073\_22302 & 3.45137 & $-$30.32072 & 10.477 & Abell 2744 & 1.2 & 27.44$\pm$0.06 & $-$19.96$\pm$0.06 & Nap25 \\
GS-20030902 & 1286\_20030902 & 53.10798 & $-$27.87760 & 10.425 & GOODS-S & -- & 27.91$\pm$0.09 & $-$19.68$\pm$0.09 & This Work \\
GS-72355 & 1286\_20030133 & 53.16594 & $-$27.83424 & 10.422 & GOODS-S & -- & 27.74$\pm$0.07 & $-$19.85$\pm$0.07 & Weib25 \\
UDS-52799 & 6368\_82699 & 34.41355 & $-$5.24425 & 10.302 & UDS & -- & 27.31$\pm$0.08 & $-$20.26$\pm$0.08 & Weib25 \\
CAPERS-COS-109917 & 6368\_109917 & 150.14291 & 2.28802 & 10.272 & COSMOS & -- & 28.00$\pm$0.11 & $-$19.57$\pm$0.11 & Tang25 \\
GHZ8 & 3073\_23984 & 3.45143 & $-$30.32180 & 10.231 & Abell 2744 & 1.2 & 26.73$\pm$0.05 & $-$20.63$\pm$0.05 & Nap25 \\
JD1$^{*}$ & 1433\_3593 & 101.98229 & 70.24327 & 10.170 & MACS 0647 & 8.0 & 25.03$\pm$0.02 & $-$20.27$\pm$0.02 & Hsiao24 \\
JD2$^{*}$ & 1433\_3349 & 101.97133 & 70.23972 & 10.170 & MACS 0647 & 5.3 & 25.51$\pm$0.03 & $-$20.23$\pm$0.03 & Hsiao24 \\
JD1c$^{*}$ & 1433\_3621 & 101.98280 & 70.24387 & 10.170 & MACS 0647 & 8.0 & 28.15$\pm$0.23 & $-$17.14$\pm$0.23 & Hsiao24 \\
JD2c$^{*}$ & 1433\_3314 & 101.96986 & 70.23939 & 10.170 & MACS 0647 & 5.3 & 27.45$\pm$0.12 & $-$18.29$\pm$0.12 & Hsiao24 \\
GHZ9 & 3073\_17724 & 3.47876 & $-$30.34552 & 10.137 & Abell 2744 & 1.3 & 27.38$\pm$0.06 & $-$19.89$\pm$0.06 & Nap25 \\
MoM-z10-1 & 5224\_144739 & 34.28225 & $-$5.14288 & 10.116 & UDS & -- & 27.49$\pm$0.12 & $-$20.05$\pm$0.12 & This Work \\
CEERS-64 & 2750\_64 & 214.92278 & 52.91153 & 10.100 & EGS & -- & 27.55$\pm$0.09 & $-$19.99$\pm$0.09 & AH23b \\
GS-z10-0 & 1210\_14177 & 53.15884 & $-$27.77349 & 10.075 & GOODS-S & -- & 29.05$\pm$0.10 & $-$18.49$\pm$0.10 & CL22 \\
GLASS-z11-17225 & 3073\_18252 & 3.50731 & $-$30.34320 & 10.066 & Abell 2744 & 1.4 & 28.01$\pm$0.09 & $-$19.16$\pm$0.09 & Nap25 \\
UNCOVER-26185 & 2561\_26185 & 3.56707 & $-$30.37786 & 10.061 & Abell 2744 & 4.1 & 27.54$\pm$0.11 & $-$18.46$\pm$0.11 & Fuj24 \\
UNCOVER-37126 & 2561\_37126 & 3.59011 & $-$30.35974 & 10.019 & Abell 2744 & 2.2 & 26.75$\pm$0.05 & $-$19.92$\pm$0.05 & Fuj24 \\
CEERS-80041 & 1345\_80041 & 214.73253 & 52.75809 & 10.010 & EGS & -- & 27.10$\pm$0.10 & $-$20.43$\pm$0.10 & AH23a \\
\bottomrule
\end{tabularx}
\end{small}
\begin{flushleft}
$^{*}$ Lensed and triply-imaged source.
\end{flushleft}
\end{table*}

We scale each of the spectra in our sample (see Table~\ref{tab:confirmations}) to its NIRCam photometry closest to a rest-frame wavelength of 1750 \AA\ (generally F200W for sources at $z<13.5$ and F277W for sources at higher redshift), to account for an overall slit loss, and correct for lensing magnification where required (adopting the magnification factors from \citealt{bergamini23} for sources in Abell 2744 and from \citealt{hsiao23} for sources in MACS 0647; see Table~\ref{tab:confirmations}). In the case of the triply-imaged MACS 0647-JD galaxy \citep{hsiao23,hsiao24}, we treat components JD1 and JD2 as part of the same source and combine them via an inverse-variance weighted mean of their magnification-corrected spectra. In the same fashion, we also combine the spectra of a second, multiply-imaged source only $\sim$3 kpc away (in the source plane) and at the same redshift, referred to as the JDc system (with lensed components JD1c and JD2c; \citealt{hsiao24}).

\begin{figure*}
\center
\includegraphics[width=\textwidth]{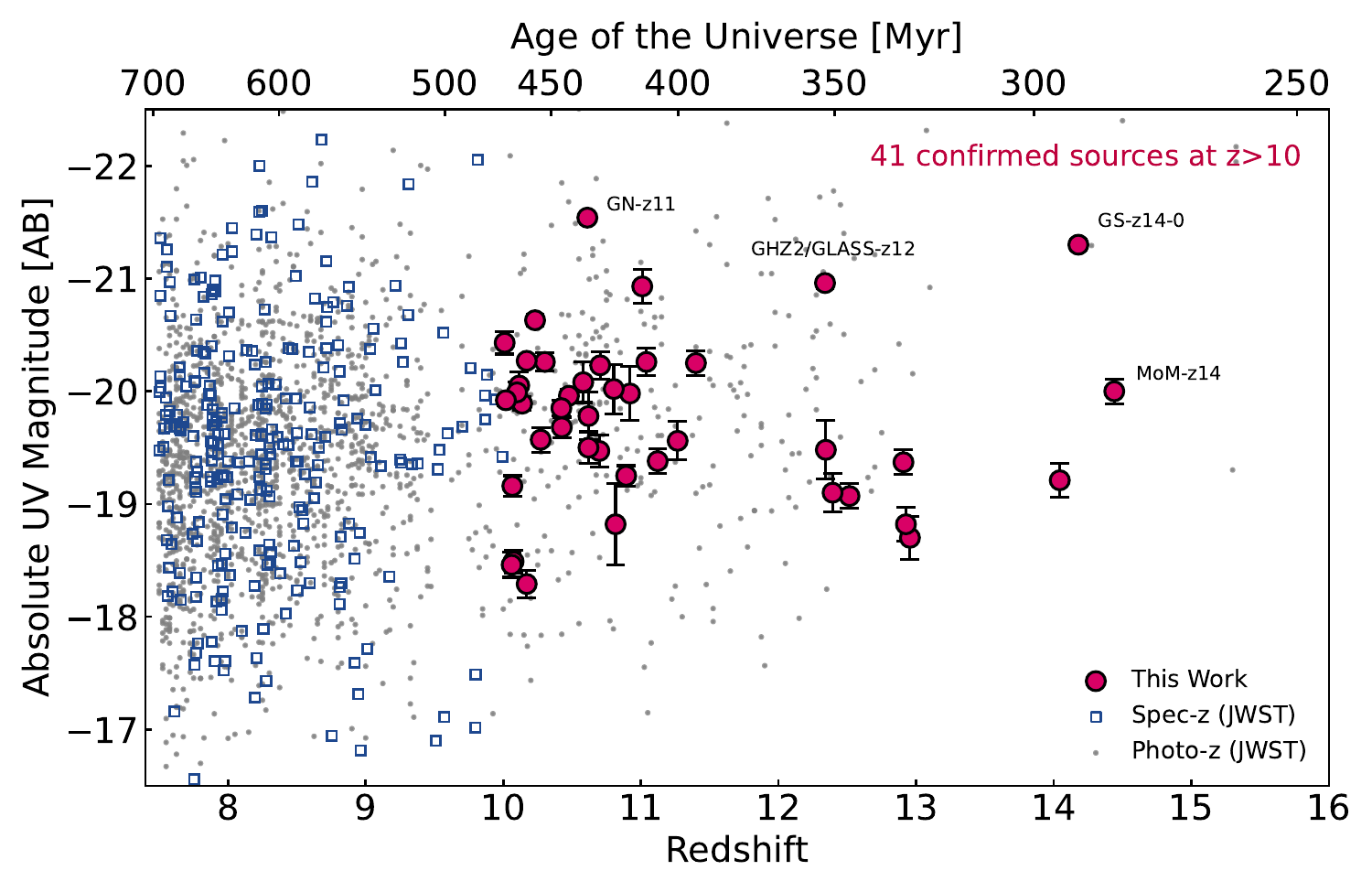}
 \caption{\textbf{Redshifts and absolute UV magnitudes of high-redshift galaxies}. Photometric objects are shown as grey points \citep{merlin24}, confirmed NIRSpec MSA sources from our extended compilation are shown as blue squares, and the $z\geqslant10$ spectroscopic sample presented in this study are shown as red circles (see Table~\ref{tab:confirmations}). A number of remarkably luminous galaxies (GN-z11 from \citealt{bunker23}, GHZ2/GLASS-z12 from \citealt{castellano24}, and JADES-GS-z14-0 from \citealt{carniani24}) and the current redshift record holder (MoM-z14 from \citealt{naidu25}) are shown for reference. In the case of the multiply-imaged MACS 0647 systems, we plot the absolute magnitudes of the brightest components (JD1 and JD2c). Spectroscopic confirmations within the first 500 Myr now number 41 unique sources from 43 spectra, when accounting for multiply-imaged lensed systems.}
 \label{fig:zspec}
\end{figure*}

In total, our procedures yield a fiducial sample of 43 spectra from 41 unique sources at $z\geqslant10$, an increase of $\sim1.4\times$ compared to the sample by \citet{tang25} and the largest of its kind thus far. We show the spectroscopic redshifts and absolute UV magnitudes of this sample in Figure~\ref{fig:zspec}, together with photometric candidates from the Astrodeep catalog for comparison. Despite the increase in numbers at such high redshifts, allowing us to move from single-source to population-level studies, the relatively modest size and heterogeneity of the sample means it will no doubt suffer from significant selection effects owing to both the challenges of identifying such distant sources as well as incompleteness from spectroscopic follow up. Moreover, given the range of absolute UV magnitudes probed here ($-22 \lesssim M_{\rm UV}\lesssim-18$ AB), the sample is characteristic only of the bright end of the UV luminosity function rather than the more representative population of early galaxies probed by intrinsically fainter sources ($M_{\rm UV}\gtrsim-17$ mag). Nonetheless, our sample represents a significant leap forward in our ability to constrain the physical properties of the earliest \textit{populations} of galaxies.

\subsection{Stacking Procedure for Composite Spectra}
\label{subsec:stacking}
In addition to measurements of individual sources, we also verify the properties of $z\geqslant10$ sources through high signal-to-noise (S/N) composite spectra constructed through stacking, allowing us measure average spectral features otherwise inaccessible in low S/N regimes. The technique has recently been used to great effect with NIRSpec spectra to determine the average properties, evolution, and emission (or absorption) line ratios of galaxies at lower redshifts (e.g., \citealt{shapley23,rb24,hu24,langeroodi24,hayes25,meyer25,glazer25,saxena25}).

\begin{figure*}
\center
\includegraphics[width=\textwidth]{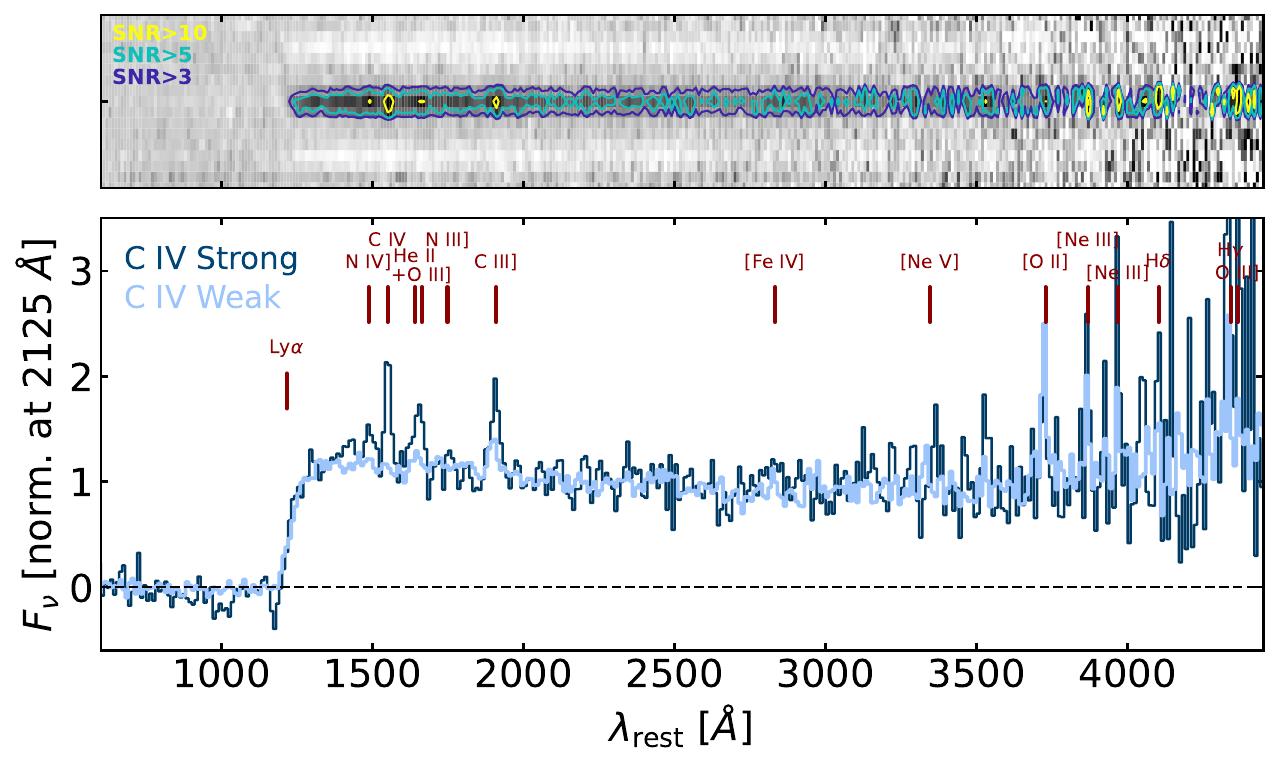}
 \caption{\textbf{Composite spectra of the \civ-strong (dark blue) and \civ-weak (light blue) samples at $z\simeq10.0-14.4$, and the corresponding 2D spectrum of the former.} Emission lines of interest (detected or not) and the Ly$\alpha$ break are indicated by red vertical lines in the 1D spectrum, as well as S/N ratio contours of 3, 5 and 10 in the 2D spectrum. The \civ-strong composite reveals strongly enhanced \niv], \civ, \heii$+$\oiii], and \ciii] line emission compared to its \civ-weak counterpart comprising the rest of the sample. The comparison highlights the strong variance across the galaxy population and diversity of astrophysical processes underpinning the evolution of galaxies within the first 500 Myr.}
 \label{fig:stack}
\end{figure*}

We employ a similar procedure to that outlined in \citet{rb24}. Briefly, all individual spectra are first de-redshifted to their rest-frame wavelengths utilizing the redshifts in Table~\ref{tab:confirmations}, then normalised to their median continuum emission at either $\lambda_{0}=2000-2250$ \AA\ (by default, or when measuring UV spectral properties) or $\lambda_{0}=3250-3500$ \AA\ (for measurements of optical line emission), and interpolated over a common wavelength grid (spanning 600-4500 \AA\ with 10 \AA\ intervals) before being added to the stack. To mitigate the effects of potential outliers from what remains a modest sample, we derive composite spectra from the median of the stack. We construct two sets of uncertainties, the first being the measurement error constructed through the propagation and averaging of the prism's error array, and the second being the standard deviation around the median through bootstrap resampling of the stacked spectra 500 times. The uncertainties of the latter are taken as the semi-difference of the 16th and 84th percentiles. Figure~\ref{fig:stack} shows two example composite spectra constructed with this procedure, one consisting of objects with identified \civ\ emission (see Section~\ref{subsec:measurements}), and one consisting of all other objects, for illustration. These composites are discussed further in Section~\ref{sec:props}.

Our normalisation procedure gives equal weighting to each of the sources and prevents a natural bias towards the most luminous objects or those benefiting from the deepest exposures, while the median ensures the composites are not affected by outliers unless they contribute meaningfully to the stack. The properties (e.g., UV continuum slopes, emission line measurements, star formation histories) of the final composites are then derived as described in Sections~\ref{subsec:measurements} and \ref{subsec:sedmethod}.

\subsection{Spectroscopic Measurements}
\label{subsec:measurements}
We measure and verify the presence of emission lines using a combination of Gaussian line fitting with an underlying power law continuum model, and visual inspection for added robustness. In cases where such lines have already been reported, we remeasure them for consistency between objects in our sample. Our adopted procedure is as follows: for each spectrum, we adopt a least-squares fit to the UV continuum using a power-law continuum model with a slope $\beta$ (where $f_{\lambda}\propto\lambda^{\beta}$), in addition to Gaussian profiles for each of the relevant lines (assuming a single Gaussian for closely-spaced doublets or individual lines, e.g., \heii$\lambda$1640 \AA\ and [\oiii]$\lambda\lambda$1660,1666 \AA). The fits are performed over two wavelength ranges, specifically $\lambda_{\rm rest}=1300-2200$ \AA\ to capture rest-frame UV lines and $\lambda_{\rm rest}=3500-4500$ \AA\ to capture rest-frame optical lines (we note that all lines redward of [\oiii]$\lambda$4363 \AA\ are generally inaccessible with NIRSpec due to the instrument's red wavelength limit and the high redshifts probed here). The central wavelengths of the lines are allowed to deviate $\pm25$ \AA\ from their expected positions to account for redshift uncertainties, while we utilise a common intrinsic line width convolved to the spectral resolution in order to minimise chances of fitting spurious peaks. All equivalent width measurements make use of the median continuum flux underneath the central position of the relevant line. The fitting process is repeated $N=1000$ times, each time perturbing the observed spectrum by its associated $1\sigma$ uncertainty array. The resulting measurements (continuum slope, line flux, rest-frame equivalent widths) and their uncertainties are chosen as the median and semi-difference of the 16th and 84th percentiles of the resulting best-fit distributions.

We measure two separate S/N ratios (SNR) to guard against noise spikes and the prism's strong wavelength-dependent spectral resolution. The first is a SNR on the equivalent width (EW) using measurement errors and the second is the peak SNR using the best-fit amplitude of the line and the standard deviation of line-free continuum flux immediately adjacent to the lines. We deem a line detected if a \textit{total} SNR (the quadratic sum of the two measurements) reaches $>2\sigma$ and satisfies visual inspection, while for non-detections we use the $1\sigma$ uncertainty from the best-fit as an upper limit. For sources where lines are identified, we refine their spectroscopic redshifts through those lines. For reference, the uncertainties associated with single-object measurements come from the measurement errors, while those of composite spectra come from bootstrapped errors.

\begin{figure}
\center
\includegraphics[width=\columnwidth]{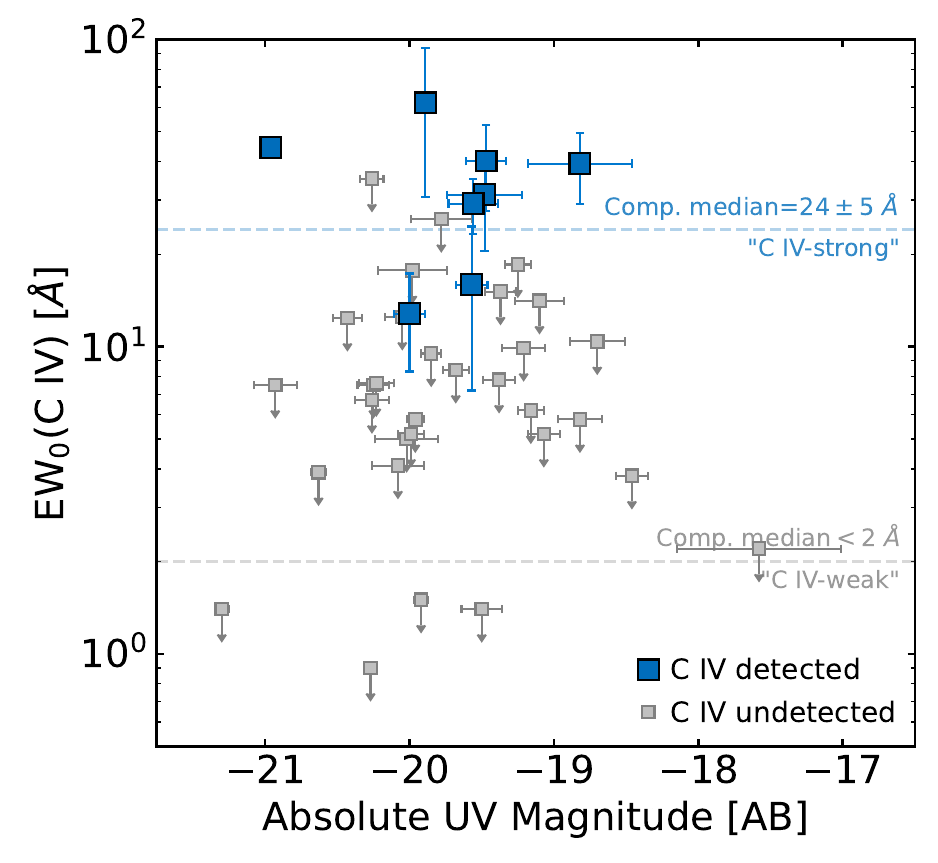}
 \caption{\textbf{Detections of \civ\ in luminous $z\geqslant10$ galaxies}. Sources with \civ\ detections (the ``\civ-strong'' sample) and those without (the ``\civ-weak'' sample) are shown as blue squares and grey upper limits, respectively. The median EWs measured on the composite spectra of the two-subsamples shown in Figure~\ref{fig:stack} are highlighted as dashed lines (where the grey line represents an upper limit), validating the individual upper limits as physical.}
 \label{fig:ewratio}
\end{figure}

In total, we identify 19 sources with detected \ciii] line emission and 8 sources with detected \civ\ emission (MoM-z14, CAPERS-EGS-65480, GHZ2/GLASS-z12, JADES-GS-20015720, CAPERS-EGS-43539, GHZ4, CAPERS-COS-109917, and GHZ9), and showcase these as a function of their \civ\ EWs in Figure~\ref{fig:ewratio}. The measured values are also reported in Table~\ref{tab:line_ews}.

\begin{table}
\centering
\caption{Rest-frame \ciii]\ and \civ\ line equivalent widths for the sources reported in Table~\ref{tab:confirmations}.}
\label{tab:line_ews}
\begin{small}
\begin{tabular*}{\columnwidth}{@{\extracolsep{\fill}}lcc}
\toprule
Name & EW$_{0}$(\ciii]) & EW$_{0}$(\civ) \\
     & [\AA] & [\AA] \\
\midrule
MoM-z14 & 16.3$\pm$7.7 & 12.8$\pm$4.5 \\
GS-z14-0 & 8.8$\pm$3.4 & $<$1.4 \\
GS-z14-1 & $<$10.4 & $<$9.9 \\
JADES-GS-z13-1-LA & $<$4.2 & $<$10.4 \\
GS-z13-0 & $<$6.2 & $<$5.8 \\
UNCOVER-13077 & 32.2$\pm$24.2 & $<$15.1 \\
GS-z12-0 & 27.5$\pm$7.8 & $<$5.2 \\
UNCOVER-38766 & 33.8$\pm$30.4 & $<$14.1 \\
CAPERS-EGS-65480 & $<$6.5 & 31.2$\pm$10.7 \\
GHZ2 & 21.1$\pm$2.2 & 44.5$\pm$2.5 \\
CEERS-1 & 22.6$\pm$12.1 & $<$7.5 \\
JADES-GS-20015720 & 19.8$\pm$5.5 & 29.2$\pm$5.9 \\
GS-z11-0 & 9.5$\pm$6.1 & $<$7.8 \\
CEERS-10 & $<$10.2 & $<$6.7 \\
CAPERS-UDS-z11 & $<$8.9 & $<$7.5 \\
MoM-z11-1 & $<$12.7 & $<$17.7 \\
JADES-GS-20177294 & 48.7$\pm$32.1 & $<$18.5 \\
CAPERS-EGS-43539 & $<$1.4 & 39.4$\pm$10.2 \\
MoM-z11-2 & 18.2$\pm$10.1 & $<$5.0 \\
EGS-22637 & $<$9.4 & $<$7.6 \\
GHZ4 & 19.4$\pm$14.7 & 40.2$\pm$12.6 \\
JADES-GS-20176151 & $<$2.8 & $<$1.4 \\
EGS-69 & $<$34.7 & $<$26.0 \\
GN-z11 & 16.9$\pm$1.6 & $<$0.1 \\
CAPERS-UDS-z10 & $<$1.6 & $<$4.1 \\
GHZ7 & 19.0$\pm$10.4 & $<$5.8 \\
GS-20030902 & $<$11.9 & $<$8.4 \\
GS-72355 & $<$32.8 & $<$9.5 \\
UDS-52799 & $<$39.9 & $<$35.2 \\
CAPERS-COS-109917 & 27.2$\pm$11.2 & 15.9$\pm$8.7 \\
GHZ8 & $<$6.2 & $<$3.9 \\
JD & 8.1$\pm$2.8 & $<$0.9 \\
JDc & $<$16.1 & $<$4.8 \\
GHZ9 & 50.6$\pm$32.6 & 62.1$\pm$31.4 \\
MoM-z10-1 & $<$1.2 & $<$12.5 \\
CEERS-64 & $<$6.8 & $<$5.2 \\
GS-z10-0 & $<$4.6 & $<$2.2 \\
GLASS-z11-17225 & $<$24.4 & $<$6.2 \\
UNCOVER-26185 & 22.0$\pm$13.5 & $<$3.8 \\
UNCOVER-37126 & 37.4$\pm$14.0 & $<$1.5 \\
CEERS-80041 & $<$13.0 & $<$12.4 \\
\bottomrule
\end{tabular*}
\end{small}
\end{table}

Rather than using a strict EW separation defined by uncertain \textit{individual} upper limits, here we instead classify objects with detected \civ\ emission \textit{in their prism spectra} as ``\civ-strong'', and those without as ``\civ-weak''. We note that such a selection is not a strict or complete one given the generally modest depths of the observations and the limited spectral resolution of the prism, nor is it likely to be 100\% pure. However it is evident from the resulting composites in Figure~\ref{fig:stack} and their measured \civ\ EWs, that the upper limits characterising the \civ-weak sample are meaningful. For reference, the \civ\ EWs measured from the composite spectra of the two sub-samples are $24\pm5$ \AA\ (\civ-strong) and $<2$ \AA\ (\civ-weak), validating the upper limits as physical, as well as the chosen separation criteria.

\subsection{Spectral Energy Distribution Modelling}
\label{subsec:sedmethod}
We use the \texttt{Bagpipes} \citep{carnall18} SED-fitting code to fit the NIRSpec prism spectra and gain complementary insight into the state of the underlying gas and stellar populations. We utilise a custom version of the code outlined in \citet{giovinazzo25}, modified to include a picket-fence model that accounts for the escape fraction of UV ionising photons, together with an updated CLOUDY grid that accounts for extreme ionising parameters up to log\,$U=$0 and hydrogen densities up to 10$^{3}$ cm$^{-3}$. We assume a non-parameteric star formation history (SFH) with a bursty continuity prior \citep{leja19,tacchella22}, which includes seven age bins with 3, 5, 10, 50, 100, 200 Myr, and the maximum age allowed by the galaxy's redshift. The continuity component of the prior captures smooth, extended star formation histories, while the bursty component enables episodic, burst-dominated SFHs over shorter timescales. The spectra are modelled with the BPASS version 2.2.1 stellar population models (including binary stellar evolution) of \citet{stanway18}, a \citet{salim18} SMC-like dust law, and the broken power law initial mass function (IMF) of \citet{stanway18}. Nebular continuum emission is also included in the model. The free parameters are thus each of the star formation bins, the stellar mass ($M_{*}$), metallicity ($Z$, assumed to be the same for gas and stars), ionisation parameter (log\,$U$), escape fraction ($f_{\rm esc}$), and dust attenuation ($A_{\rm v}$). The best-fit values are taken as the median of the posterior and the semi-difference of the 16th and 84th percentiles, respectively. We refer interested readers to \citet{giovinazzo25} for details on the modifications and modelling procedures.

\subsection{Modelling Intrinsic Scatter}
\label{subsec:emcee}
Of our primary goals (determining the prevalence of anomalous UV line emitters, characterising the physical conditions driving the distribution of their spectral and morphological features, determining the timescales on which these occur), one is to determine the underlying distribution and intrinsic scatter of galaxy observables, in order to assess the global properties driving spectral diversity. To do this, we model the distribution and intrinsic scatter of our measurements using a censored likelihood formalism that incorporates both detections and upper limits. For detections, where a measured value and associated uncertainty are available, the likelihood is evaluated under a specified probability distribution (e.g., normal or log-normal) that accounts for both measurement error and intrinsic population scatter (i.e., $\sigma_{\rm tot}=\sqrt{\sigma^{2}_{\rm int} + \sigma^{2}_{\rm err}}$). For non-detections, upper limits are included via the cumulative probability that the true value lies below the detection threshold, given the same underlying model. This approach ensures that both detections and censored observations inform the inferred distribution.

With the above in mind, to evaluate the best-fit parameters (i.e., the median of a probability distribution and its intrinsic scatter), we use the affine-invariant MCMC sampler, \texttt{emcee} \citep{emcee}, to draw samples from the posterior distribution, combining our likelihood function with flat priors. We use 500 walkers with 500 steps and discard the first 100 as ``burn-in'' steps. In all cases we adopt the median and 16th-84th percentile range as our fiducial values (the latter to account for potential asymmetries in the posterior).

\section{The Distribution and Intrinsic Scatter of $\lowercase{z}\geqslant10$ Galaxy Observables}
\label{sec:distribution}
The diversity of the $z\geqslant10$ population can be explored through the distribution and intrinsic scatter of spectral and morphological features. Here we focus on those accessible to NIRSpec and NIRCam for sources at $z\geqslant10$, specifically the UV line strengths, UV continuum slopes, and half-light radii. To first order, unimodal distributions would suggest shared global properties deriving from a common evolutionary pathway, while bimodal (or multimodal) distributions would suggest distinct populations. Applying this logic to our sample, in the following sections we characterise the distributions and scatter of the \civ-weak population, using measurements available from the NIRSpec and NIRCam data, and compare these to measurements of the \civ-strong population in order to determine the latter's place relative to the more ``typical'' population. The resulting distributions and their best-fit parameters are summarised in Table~\ref{tab:distributions}.

\begin{table}
\centering
\caption{Best-fit parameters (median and intrinsic scatter) of the normal and log-normal parametrisations to the observables for the \civ-weak population discussed in Section~\ref{sec:distribution}. The numbers on each parameter represent the median, 16th, and 84th percentiles of the posterior. The median $\mu$ is given in linear units, while the $1\sigma$ scatter is given in natural log units (for log-normal functions). Note that the compact and extended $r_{\rm 50}$ distributions are modelled simultaneously.}
\label{tab:distributions}
\begin{tabular}{lccc}
\toprule
Observable & Function & $\mu$ & $\sigma_{\rm ln,int}$ \\
\midrule
\ciii] EW [\AA]           & log-normal & $8.3^{+2.9}_{-2.6}$          & $2.5^{+1.3}_{-0.7}$ \\
$\beta$                   & normal     & $-2.23^{+0.03}_{-0.03}$      & $0.11^{+0.04}_{-0.03}$ \\
$r_{\rm 50}$ (compact) [pc]  & log-normal & $70.4^{+278.0}_{-44.1}$     & $1.4^{+0.9}_{-0.7}$ \\
$r_{\rm 50}$ (extended) [pc] & log-normal & $459.1^{+127.1}_{-324.8}$   & $0.7^{+0.7}_{-0.3}$ \\
\bottomrule
\end{tabular}
\end{table}

\subsection{High-Ionisation \ciii] EWs}
\label{subsec:ciii}
\begin{figure}
\center
\includegraphics[width=\columnwidth]{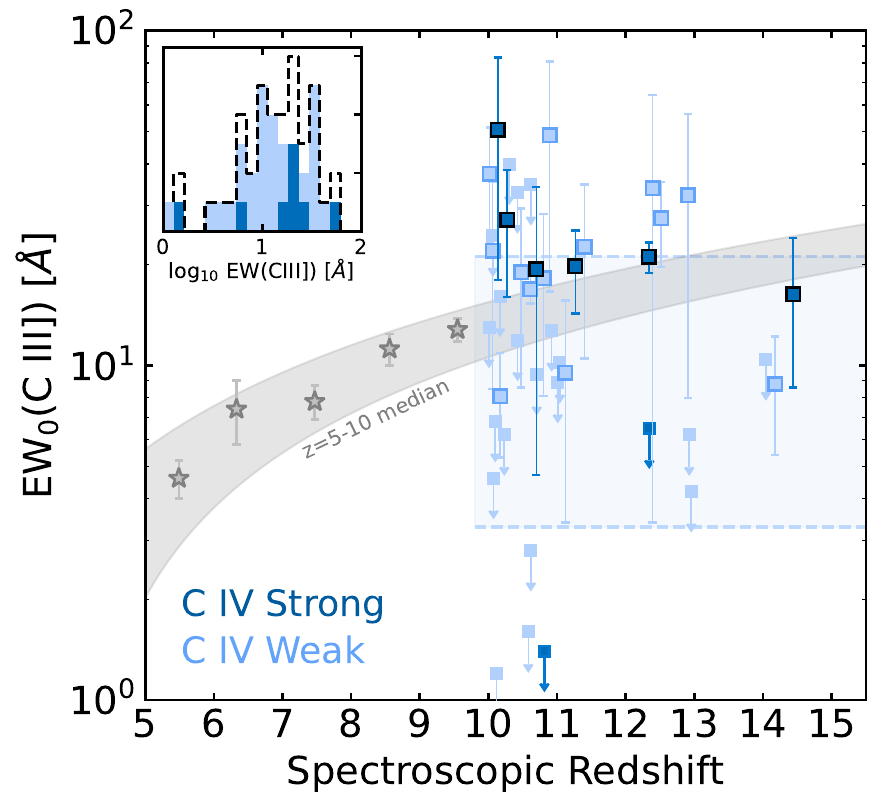}
 \caption{
 \textbf{The \ciii] rest-frame equivalent widths of high-redshift galaxies.} Detections and $1\sigma$ upper limits for our fiducial $z\geqslant10$ sample are shown as light (\civ-weak) and dark (\civ-strong) blue squares and arrows, respectively. The intrinsic scatter around the median of the \civ-weak EW distribution (including detections and upper limits) is shown as a light-blue filled box, while the inset axis shows a histogram of the full sample adopting the same color scheme (in addition to the combined sample with a black dashed line). The median values of $z=5-10$ galaxies and an extrapolation of their evolution, from the composite spectra of \citet{rb24}, are shown as grey stars and fill, respectively.}
 \label{fig:ciii}
\end{figure}

\ciii]$\lambda\lambda$1907,1909 \AA\ emission in the rest-frame UV serves as a standout tracer for star formation in metal-poor and high-ionisation environments \citep{stark17,berg19a}. As such, EWs of the line represent a key indicator for variations in the recent star formation activity, metallicities, and ionisation parameters of $z\geqslant10$ galaxies when strong [\oiii]+H$\beta$ and H$\alpha$ lines are lacking. We identify prominent \ciii] emission across a significant fraction of the sample (19/41 sources, or $\sim$46\%) and plot the resulting EWs as a function of redshift in Figure~\ref{fig:ciii}. As expected given the resolution and depths of the prism spectra, only the strongest emission is identified in individual sources, spanning equivalent widths 8-51 \AA\ across redshifts 10 to 14.4. While a small number of upper limits populate zones of especially high equivalent width ($>$15 \AA, 5 sources), the majority trace lower strengths of 1-15 \AA\ (17 sources), suggesting low star formation activity. In comparison, all but two \civ-strong sources reside in regions with EW$_{0}$(\ciii])$\gtrsim15$ \AA.
We find the distribution of \ciii] EWs (accounting for detections and upper limits) from \civ-weak sources to be well parametrised by a log-normal with a median of $\mu=8.3^{+2.9}_{-2.6}$ \AA\ and intrinsic scatter of $\sigma_{\rm ln}=2.5^{+1.3}_{-0.7}$ \AA, albeit with a tail towards a lower median and towards higher scatter driven by the significant number of upper limits. The intrinsic scatter around the median is shown in Figure~\ref{fig:ciii}, within which we find all but one of the strong \civ\ emitters fall inside (within uncertainties) and consistent with the high-end range of EWs set by the \civ-weak objects. In other words, we find no evidence to suggest \civ-strong objects show particularly enhanced or reduced \ciii] EWs compared to the extended range set by the \civ-weak sample. We explore the possible origins of the scatter within this EW range in Section~\ref{subsec:metalpoor}.

A number of recent JWST-led studies have sought to characterise the cadence and evolution of \ciii] emission in $z\simeq5-10$ galaxies (e.g., \citealt{rb24,hayes25,deugenio23,carniani24,tang25}), finding a high prevalence in stacked spectra and increasing EWs with redshift. Of particular interest is the shape of such evolution; at redshifts $z\simeq5-10$, \citet{rb24} found a clear linear evolution with $d$(EW)/d$z$$\sim$1.84 \AA, suggesting increasingly hard ionisation fields and declining metallicities. The enlarged sample of $z\geqslant10$ detections measured here is consistent with previous measurements and appears to continue the trend reported by \citet{rb24}, at least to $z\sim12$, with some suggestion of a possible downturn at redshifts beyond 13 as traced by the three especially luminous sources at $z\sim14$ with reduced EWs.

\subsection{UV Continuum Slopes}
\label{subsec:beta}
The UV continuum slope ($\beta$) in high-redshift galaxies is regulated primarily by the attenuation of the continuum by dust, and/or by the contributions of young and hot O- and B-type stars. Thus, although secondary effects also play a role, the distribution of $\beta$ primarily informs the build up of dust grains and the stellar ages of high redshift sources, each representing crucial constraints in theoretical models of early galaxy evolution (e.g., \citealt{ferrara23,mason23}). Recent works have sought to characterise $\beta$ across high-redshift samples with both photometric (e.g., \citealt{cullen23,cullen24,topping24a,austin24}) and spectroscopic (e.g., \citealt{rb24,dottorini24,saxena25,donnan25,tang25}) data sets, finding a notable evolution towards bluer slopes at higher redshift that flattens out beyond $z\simeq7$, albeit with significant scatter. The implication of the measurements is a rapid shift towards dust-poor or dust-free systems \citep{mauerhofer25}, however, extrapolations of the trend beyond $z\simeq10$ remain largely spectroscopically unverified.

\begin{figure}
\center
\includegraphics[width=\columnwidth]{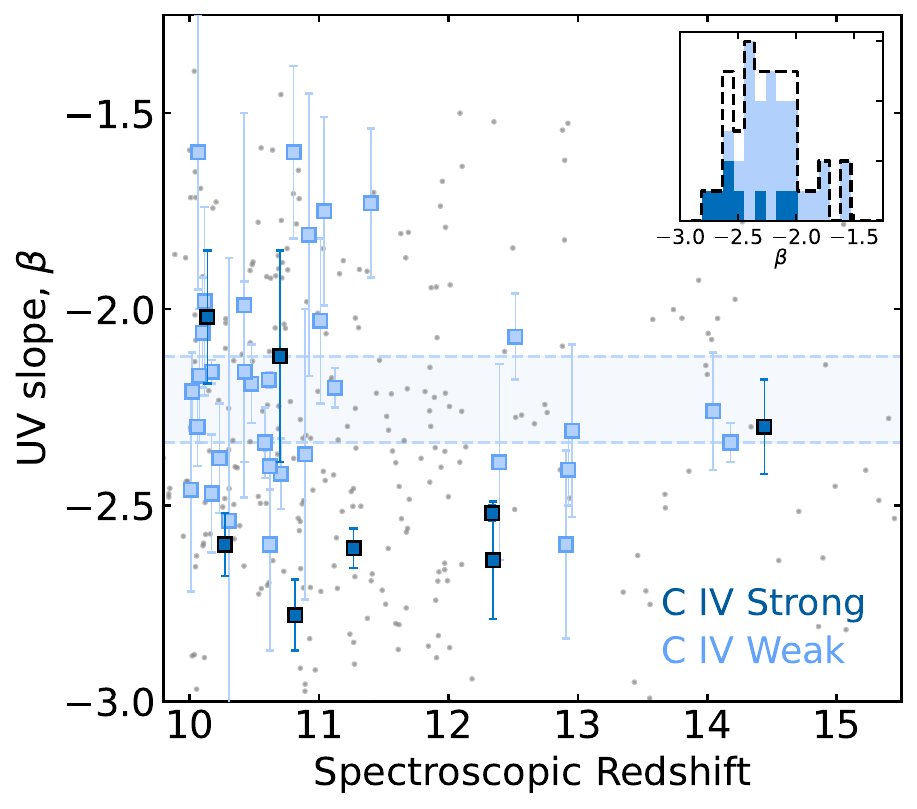}
 \caption{
 \textbf{The UV continuum slopes of $z\geqslant10$ galaxies with redshift}. Points adopt the same color scheme as in previous figures. The intrinsic scatter around the median (of the \civ-weak population) is plotted as a light-blue box, while a histogram of the distribution in shown in the inset axes (black line denoting the combined sample of \civ-weak and \civ-strong objects). The range of observed $\beta$ values suggests the bulk of the population are dust-poor or dust-free. The photometric measurements of \citet{weibel25} are shown in grey for comparison.}
 \label{fig:beta}
\end{figure}

The UV slopes resulting from our fitting procedure are shown in Figure~\ref{fig:beta}. We find the bulk of our measurements trace a relatively limited range of UV slopes between $\beta\approx-2$ and $\beta\approx-2.5$ (28 objects, 68\% of the sample), with a small number of outliers displaying either especially red slopes ($\beta\simeq-1.50$ to $-1.75$, five objects) or very blue slopes ($\beta\simeq-2.50$ to $-2.75$, eight objects). The relatively narrow range from the bulk of the sample shows no noticeable evolution over the $\sim200$ Myr probed, perhaps suggesting the majority of this population remains predominantly dust-free (or extremely dust-poor) given the short timescales to build up a significant reservoir of dust grains around their stars. Although the sample size remains modest, the lack of clear evolution is mirrored in the larger photometric sample of \citet{weibel25} and its scatter. The median and intrinsic scatter of the \civ-weak $\beta$ distribution (assuming a normal distribution) is measured as $\mu=-2.23^{+0.03}_{-0.03}$ and $\sigma=0.11^{+0.04}_{-0.03}$, respectively, which we highlight in Figure~\ref{fig:beta}. The scatter appears small, and virtually all \civ-weak sources are consistent with the median within uncertainties. The only exceptions are the \civ-weak sources with especially red slopes ($\beta>-2.0$), and the \civ-strong sample with especially blue slopes ($\beta<-2.5$), both of which appear as outliers to the scatter at $>3\sigma$. Although some of this scatter may be impacted by noise, it is notable that most (5/8 objects) of the \civ-strong objects are clustered at the bluest end of the distribution, yet not separate from the \civ-weak distribution. This points to ISM and stellar conditions that only a small fraction of the \civ-weak objects may be subject to.
The overall blue slopes traced by the combined populations are consistent with values expected from young stellar ages, low dust content, and mild reddening by the nebular continuum \citep{katz24,saxena25,rb24}. We discuss a more detailed physical description of our $\beta$ measurements and the scatter in Section~\ref{subsec:betaorigins}.


\subsection{Half-Light Radii}
\label{subsec:morphologies}
The physical sizes of galaxies are the natural consequence of the physics governing their early evolution, and as such serve as a useful indicator for discerning sub-populations of galaxies. As an illustration of this, a number of recent works have shown a tentative size dichotomy between $z\geqslant10$ sources displaying strong \niv] emission and their \niv]-weak counterparts, with the former displaying compact sizes of $\lesssim100$ pc and the latter more extended sizes of $\sim200-500$ pc \citep{harikane25,naidu25}. Such a dichotomy could hint at different evolutionary processes, however comparisons have thus far remained exclusive to only a small handful of objects. With an enlarged sample in hand, here we seek to verify the presence (or otherwise) of such a dichotomy.

\begin{figure}
\center
\includegraphics[width=\columnwidth]{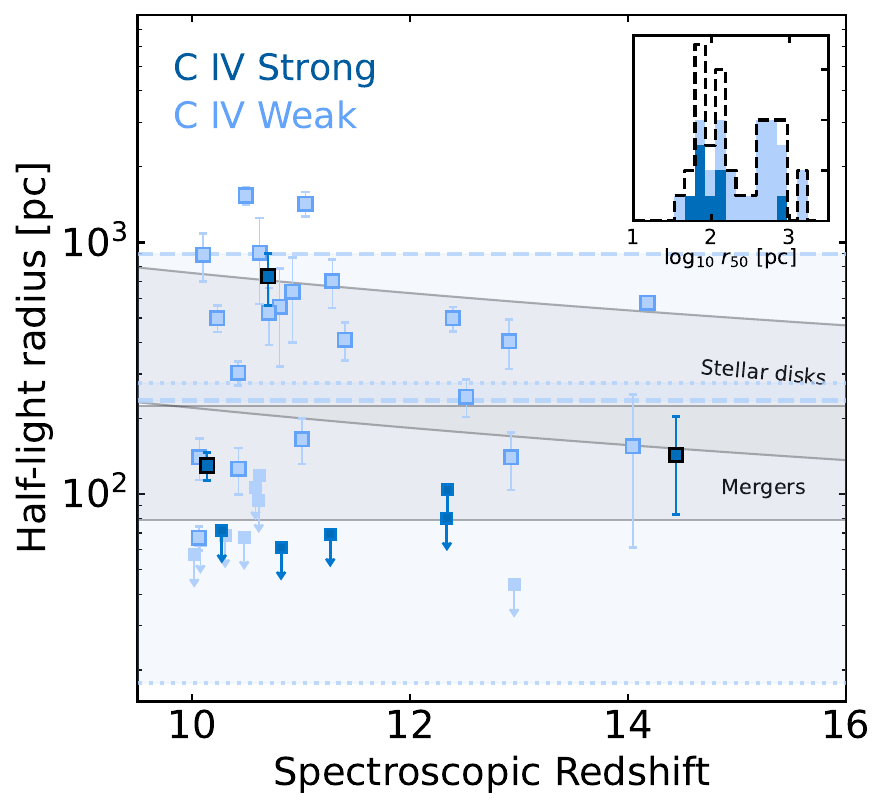}
 \caption{\textbf{Half-light UV radii of $z\geqslant10$ galaxies.} 
 The wide range of radii measured from NIRCam/F200W imaging showcases the complex morphologies at play, consistent with both merger-dominated (compact) and disk-dominated (extended) interpretations. Points adopt the same color scheme as in previous figures, while the intrinsic scatter of our log-normal parametrisations are shown as light-blue boxes. Upper limits indicate sources with half-light radii less than 0.75 pixels. A histogram of the points and the full sample is shown in the inset axis. The log-normal size distributions from \citet{ono25} (mergers) and \citet{shibuya15} (stellar disks), where the timing relative to a merger in compact sources and outshining effects in extended sources regulate the observed scatter, are shown as grey-filled regions. Strong \civ\ emitters occupy almost exclusively the parameter space tracing very compact objects, albeit with overlap with some \civ-weak sources.}
 \label{fig:morpho}
\end{figure}

We determine UV-based half-light radii using PySersic \citep{pasha23} on 0.04 arcsec/pixel resolution (corresponding to $\sim170$ pc at $z=10$) NIRCam F200W mosaics from the DJA and plot the resulting values in Figure~\ref{fig:morpho}. We note that some of our sources are lensed, although the magnification is generally modest ($\mu\simeq1.2-2.0$) except for the triply-imaged source MACS0647-JD with $\mu\simeq5.3-8.0$. We exclude this source from this analysis and warn that our circularised estimates should be treated as first-order estimates. A non-negligible fraction ($34$\%) of the remaining sources have half-light radii $<1$ pixel, indicating these sources are unresolved relative to the point spread function (PSF). With a full width at half maximum (FWHM) of $\sim0.06''$ (or 1.5 pixels; \citealt{weibel25}), we deem a source unresolved if it is characterised by a radius less than 0.75 pixels, and treat this as an upper limit at its inferred value given the semi-arbitrary nature of the threshold.

The sample clearly exhibits a large range of radii, $r_{\rm 50}\simeq40-1000$ pc, and highlights diverse structure in the early Universe. The range of sizes is consistent with those inferred from photometric samples (e.g., \citealt{shibuya15,morishita24,ono25,yang25}), while the shape of the underlying distribution shows evidence for a bimodality at $10<z<12$, with two peaks at $r_{\rm 50}\simeq100$ pc and $r_{\rm 50}\simeq500$ pc. Although this is likely due in part to an incomplete sample, we nonetheless characterise each peak separately. The two peaks of the \civ-weak objects are well described by overlapping log-normal distributions with $\mu_{\rm compact}=70^{+278.0}_{-44.1}$ pc and $\mu_{\rm extended}=459.1^{+127.1}_{-324.8}$ pc, and intrinsic scatter $\sigma_{\rm ln,compact}=1.4^{+0.9}_{-0.7}$ pc and $\sigma_{\rm ln,extended}=0.7^{+0.7}_{-0.3}$ pc. The distribution of \civ-strong objects relative to these peaks also appears to validate the size dichotomy of extreme UV line emitters seen in recent works, with all but one \civ-strong source characterised by half-light radii $\lesssim100-150$ pc that are predominantly unresolved but still within the intrinsic scatter set by their \civ-weak counterparts. 

While the bimodal nature of the distribution requires verification with larger samples of confirmed sources, the large range of sizes shown here most likely reflects the interplay between a combination of dynamical processes (e.g., star formation, mergers, and stellar feedback). Indeed, each peak of the bimodal distribution is consistent with extrapolations of separate photometric UV size measurements governed by either merger or star formation induced scatter. For example, based on NIRCam observations of $z\sim10-16$ galaxies, \citet{ono25} argue for merger-induced interactions as the main driver of the scatter within their log-normal distribution tracing compact ($r_{50}\lesssim200$ pc) sources, where an enhancement or reduction in size is due to the timing relative to the merger event: sources experiencing a pre-merger phase display more extended sizes due to the temporary accumulation of stellar content, before angular momentum is lost and their sizes rapidly shrink into a post-merger phase characterised by high star formation rate surface densities with increased star formation efficiencies \citep{yajima22}. The second peak of more extended sizes is well-matched instead to extrapolations of the $0<z<10$ size evolution from \citet{shibuya15}, as well as those of \citet{yang25} and \citet{morishita24}. Here the extended morphologies are thought to be dominated by stellar-disks, where the observed scatter in size may be driven by relatively simple outshining effects such that a concentrated boost in star formation (e.g., from infalling gas) temporarily reduces the \textit{observed} galaxy size relative to its underlying disk before returning to more representative levels of star formation (e.g., \citealt{mcclymont25b}).

With this in mind, the near-ubiquity of the \civ-strong population at the most compact end of the size distribution suggests these sources may represent the clearest examples of outshining by intense bursts of star formation in dense gas clouds and/or merging activity. A relevant example of such diverse processes and morphologies, and their potential importance, is the analysis of a magnified source at $z\simeq6$ by \citet{fujimoto24}, the ``cosmic grapes''. Thanks to gravitational lensing, 15 individual star-bursting clumps with sizes $\simeq10-60$ pc contributing 70\% of the total UV luminosity were identified within a more extended disk component with a half-light radius $\simeq680$ pc. A number of further and similar examples has revealed evidence for bound star clusters in $z\sim6-10$ galaxies \citep{vanzella23,vanzella25,fujimoto24,adamo24,claeyssens25}, highlighting concentrated modes of star formation. Whether these clumps represent bound star clusters or relics from merging sources remains uncertain. However, the examples illustrate how mergers and disk instabilities can temporarily regulate observed galaxy sizes and luminosities, and the location of \civ-strong objects at the most compact end of the distribution is suggestive of temporarily enhanced star formation which possibly outshines a more extended underlying structure. Deeper and higher resolution imaging data will be required to discern the morphological origins of such strong UV line emitters.

\section{The Physical Properties of Early Galaxies}
\label{sec:props}
Having characterised the distribution and scatter of the observable features within our 
sample, we now explore what these imply for the physical properties of galaxies observed within the first 500 Myr. Specifically, we aim to evaluate whether those of strong \civ\ emitters differ significantly from the rest of the sample which would suggest an alternate physical framework (e.g., exotic stellar populations, heavy-seed supermassive black holes, enhanced star formation efficiencies, top-heavy and/or evolving IMF), or whether they simply represent the tail of the overall distribution consistent with some evolutionary cycle of activity. To enable robust comparisons, we analyse composite spectra of the \civ-weak population constructed from three bins for each set of measurements from Section~\ref{sec:distribution}, thus allowing us to assess correlations between their spectral features and the scatter. For a representative \civ-strong spectrum, we use the \civ-strong composite shown in Figure~\ref{fig:stack}. The properties inferred for each of our derived composites are presented in Table~\ref{tab:stacks}.

\begin{table*}
\centering
\caption{The spectroscopic and global properties of composite $z\geqslant10$ spectra derived in this study.}
\label{tab:stacks}
\begin{small}
\begin{tabular}{lcccccccccc}
\toprule
Name & Bin Edges & $N_{\rm gals}$ & $z_{\rm spec}$ & $M_{\rm UV}$ & $\beta$ & \ciii] EW & log\,$U^{\ddagger}$ & $Z^{\ddagger}$ & Age$^{\ddagger}$ & $f_{\rm burst}^{\ddagger}$ \\
 & & & & [AB] & & [\AA] & & [$Z_{\odot}$] & [Myr] & [$<3$ Myr]\\
\midrule
\civ-strong$^{\dagger}$ & \civ-detected    & 8  & 11.04$\pm$1.34 & $-$19.56$\pm$0.57 & $-$2.39$\pm$0.06 & 23$\pm$9  & $-$0.08$\pm$0.08 & 0.09$\pm$0.03 & 16$\pm$5   & 0.69 \\
\civ-weak$^{\dagger}$   & \civ-undetected  & 33 & 10.61$\pm$1.18 & $-$19.94$\pm$0.86 & $-$2.15$\pm$0.02 & 11$\pm$2  & $-$0.40$\pm$0.26 & 0.09$\pm$0.02 & 13$\pm$5   & 0.05 \\
\midrule
\ciii]-weak      & $<10$ \AA\       & 13 & 10.66$\pm$1.30 & $-$20.03$\pm$1.00 & $-$2.20$\pm$0.03 & $<$2      & $-$2.34$\pm$1.04 & 0.01$\pm$0.01 & 30$\pm$2   & 0.03 \\
\ciii]-moderate  & 10--20 \AA\      & 9  & 10.54$\pm$1.12 & $-$20.02$\pm$0.77 & $-$2.06$\pm$0.05 & 9$\pm$5   & $-$2.22$\pm$1.15 & 0.02$\pm$0.01 & 56$\pm$4   & 0.01 \\
\ciii]-strong    & $>20$ \AA\       & 11 & 10.49$\pm$1.04 & $-$19.37$\pm$0.53 & $-$2.19$\pm$0.08 & 28$\pm$22 & $-$1.75$\pm$0.30 & 0.08$\pm$0.02 & 181$\pm$7  & 0.42 \\
\midrule
$\beta$-red      & $>-2.00$         & 7  & 10.86$\pm$0.43 & $-$20.00$\pm$0.37 & $-$1.74$\pm$0.17 & 13$\pm$7  & $-$1.38$\pm$0.17 & 0.16$\pm$0.04 & 222$\pm$18 & 0.29 \\
$\beta$-moderate & $-2.40$ to $-2.00$ & 18 & 10.49$\pm$1.34 & $-$19.92$\pm$1.04 & $-$2.18$\pm$0.02 & 7$\pm$3   & $-$0.83$\pm$0.25 & 0.09$\pm$0.02 & 22$\pm$1   & 0.18 \\
$\beta$-blue     & $<-2.40$         & 8  & 10.62$\pm$1.11 & $-$19.78$\pm$0.54 & $-$2.45$\pm$0.09 & 12$\pm$6  & $-$2.62$\pm$0.40 & 0.05$\pm$0.02 & 11$\pm$5   & 0.05 \\
\midrule
$r_{\rm50}$-compact$^{*}$  & $<100$ pc     & 7  & 10.46$\pm$1.15 & $-$19.92$\pm$1.37 & $-$2.21$\pm$0.21 & 6$\pm$5   & $-$0.96$\pm$0.46 & 0.17$\pm$0.06 & 83$\pm$16  & 0.33 \\
$r_{\rm50}$-moderate$^{*}$ & 100--400 pc   & 11 & 10.62$\pm$1.33 & $-$19.44$\pm$0.47 & $-$2.22$\pm$0.04 & 8$\pm$5   & $-$2.67$\pm$0.66 & 0.01$\pm$0.01 & 31$\pm$5   & 0.12 \\
$r_{\rm50}$-extended$^{*}$ & $>400$ pc     & 13 & 10.86$\pm$1.07 & $-$20.02$\pm$0.61 & $-$2.06$\pm$0.06 & 10$\pm$4  & $-$0.53$\pm$0.32 & 0.02$\pm$0.01 & 25$\pm$2   & 0.14 \\
\bottomrule
\end{tabular}
\end{small}

\begin{flushleft}
$^{\ddagger}$ Properties estimated from SED-fitting with \texttt{Bagpipes}.\\
$^{\dagger}$ The \civ\ EWs are 24$\pm$5 \AA\ and $<$2 \AA\ for the \civ-strong and \civ-weak composites, respectively.\\
$^{*}$ Excluding lensed sources.
\end{flushleft}
\end{table*}

\subsection{Metal-poor Star Formation and Hard Radiation Fields}
\label{subsec:metalpoor}
To verify the metal contents and star formation activity of our sample, we analyse composite spectra of the \civ-weak population binned in \ciii] EW (namely 0-10 \AA, 10-20 \AA, and 20-100 \AA, based on the 16th and 84th percentiles of the distribution). The resulting \ciii] EWs are 28$\pm$22 \AA, 9$\pm$5 \AA, and $<$2 \AA\ for the \ciii]-strong (11 sources), -moderate (9 sources), and -weak (13 sources) composites, while the \civ-strong composite has 23$\pm$9 \AA. Leveraging the higher S/N in the rest-frame optical, we derive oxygen abundances based on the Ne3O2 and RO2Ne3 calibrations of \citet{sanders25}, which result in metallicities of 12$+$log(O/H)$\simeq7.23\pm0.24$ and $7.07\pm0.39$ for the \ciii]-strong and \ciii]-moderate composites, respectively, while we infer a \texttt{Bagpipes}-based oxygen abundance of $6.72\pm0.03$ for the \ciii]-weak composite given the non-detection of rest-optical lines. Similarly, the optical calibrations yield 
12$+$log(O/H)$\simeq7.33\pm0.29$ for the \civ-strong composite. Assuming a solar metallicity of 12$+$log(O/H)$_{\odot}=8.69$ from \citet{asplund09}, these correspond to 0.034$\pm$0.019 $Z_{\odot}$, 0.024$\pm$0.022 $Z_{\odot}$, 0.011$\pm$0.001 $Z_{\odot}$, and 0.043$\pm$0.029 $Z_{\odot}$. For comparison, the \texttt{Bagpipes}-inferred metallicities for the \ciii]-strong, \ciii]-moderate, and \civ-strong composites are 0.081$\pm$0.023 $Z_{\odot}$, 0.023$\pm$0.009 $Z_{\odot}$, and 0.088$\pm$0.029 $Z_{\odot}$, in reasonable agreement within uncertainties. These values indicate that while these objects are characterised by very metal-poor ISM conditions, the sample as a whole already shows significant chemical enrichment after only a few 100 Myr of evolution and cannot be considered first-generation galaxies (c.f. with \citealt{deugenio23,cameron23_nitrogen,scholtz25}), consistent with the simulations of \citet{katz23b} who highlight the very rapid chemical enrichment of the ISM from pristine conditions.

\begin{figure}
\center
\includegraphics[width=\columnwidth]{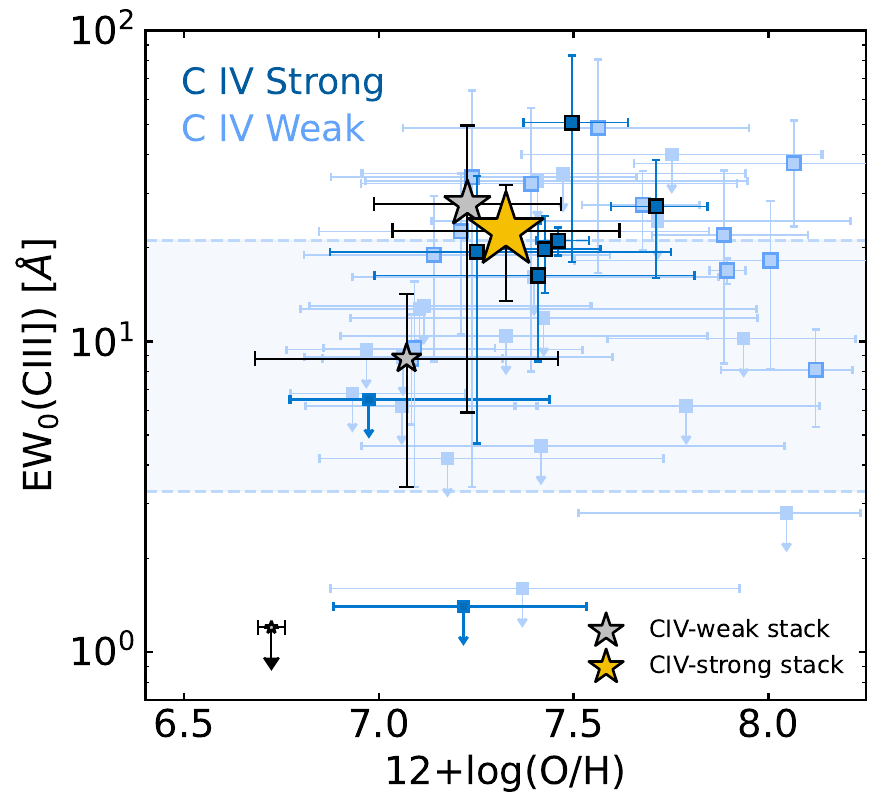}
 \caption{\textbf{The metallicities of $z\geqslant10$ galaxies.} Individual points (small squares) follow the same color scheme as in previous figures and use \texttt{Bagpipes}-inferred metallicities. Composite measurements are plotted as stars (grey for \civ-weak composites binned in \ciii] EW, and gold for the \civ-strong composite), scaled in size as a function of their recent (0-10 Myr) star formation activity with respect to the total star formation over their lifetime, where larger size indicates more recent star formation. The light-blue box shows the intrinsic scatter of \ciii] EW for the \civ-weak population measured in Section~\ref{subsec:ciii}. The size-scaling of the composites together with the limited range in metallicities suggest star formation activity is the main regulator of the \ciii] EW scatter.}
 \label{fig:metal}
\end{figure}

The spectroscopic measurements of the composites and the \texttt{Bagpipes} individual metallicities are plotted in Figure~\ref{fig:metal}, where the latter show significant scatter ($\sim1.4$ dex) between a range of metallicities 12$+$log(O/H)$\simeq6.74-8.12$ (0.011-0.270\,$Z_{\odot}$) and only modest variation with \ciii] EW. This is even more evident from the modest range of metallicities seen in the composite spectra, which display extremely limited variation ($\sim0.4$ dex, consistent with zero within uncertainties) as a function of \ciii] EW. Both \civ-weak and \civ-strong objects display similarly low metallicities, with differences of 0.1-0.5 dex (again consistent with zero, within uncertainties) between the \civ-strong and \civ-weak composites. The small differences in metallicities suggest these are unlikely to be the main driver behind the large intrinsic scatter of observed \ciii] EWs from Section~\ref{subsec:ciii}, nor the primary regulator of strong \civ\ emission.

Since star formation represents a key regulator of \ciii] emission, we also verify the ionising conditions of the composites. We find some indication of a variation in ionisation parameters from \texttt{Bagpipes}, with log\,$U\simeq-1.75\pm0.30$, $-2.22\pm1.15$, and $-2.34\pm1.04$ for the \ciii]-strong, -moderate, and -weak composites respectively, and a far more intense ionisation field of log\,$U\simeq-0.08\pm0.08$ associated with the \civ-strong composite. To verify this further, we scale the size of the composite points in Figure~\ref{fig:metal} according to the recent star formation activity estimated from \texttt{Bagpipes}, i.e., the amount of stars formed within the last 10 Myr relative to the total star formation over the full lifetime of the source. We clearly see a strong positive correlation between the recent star formation activity and the EW of \ciii], highlighting the interconnection. Noticeably, we find the \civ-strong composite shows the most significant star formation activity, likely pointing to a requirement of intense episodes of star formation in metal-poor environments to power \civ. As such, we attribute the scatter of \ciii] EWs to variations in recent star formation activity rather than vastly different metallicities.

\subsection{Dust-Poor Interstellar Media and Young Stellar Populations}
\label{subsec:betaorigins}
We aim to shed light on the main regulators of $\beta$ within the \civ-weak sample by verifying contributions from dust attenuation, varying stellar ages, and nebular continuum. We generate composite spectra of the \civ-weak population in bins of $\beta$, namely $-3.00$ to $-2.40$, $-2.40$ to $-2.00$, and $-2.00$ to $-1.50$ (again guided by the 16th and 84th percentiles of the distribution). We measure average UV slopes of $-$1.74$\pm$0.17 (7 sources), $-$2.18$\pm$0.02 (18 sources), and $-$2.43$\pm$0.09 (8 sources) for the reddest, moderately-blue, and bluest composites, respectively.

Verifying dust attenuation, we first adopt the $\beta$-$E(B-V)$ relation by \citet{reddy18} for young and metal-poor sources (100 Myr and 0.14\,$Z_{\odot}$), assuming the SMC-like dust attenuation curves of \citet{gordon03} and \citet{reddy15}. Applying the relation to both individual sources and our composites, we find all but the five reddest individual sources ($>85$\% of the sample) show $E(B-V)$ values consistent with either dust-free or dust-poor conditions ($E(B-V)\lesssim0.00-0.13$), within uncertainties. Those remaining five sources (CEERS-1, CEERS-10, MoM-z11-1, MoM-z11-2, and GLASS-z11-17225) display slightly redder values of $E(B-V)\simeq0.07-0.22$, indicative of moderately enhanced dust attenuation. We note the \citet{gordon03} curve yields a slightly tighter distribution (for the full sample) with $E(B-V)\simeq0.00-0.09$ that favours more pristine conditions, compared to that of \citet{reddy15} which yields $E(B-V)\simeq0.00-0.22$. The same conclusion holds for the composite spectra, for which we infer color excesses of $E(B-V)$=0.077$\pm$0.016 (0.189$\pm$0.039), 0.039$\pm$0.002 (0.095$\pm$0.004), and 0.017$\pm$0.008 (0.040$\pm$0.020) for the reddest, moderately-blue, and bluest composites respectively, assuming the \citet{gordon03} \citep{reddy15} extinction curve. For comparison, the \civ-strong composite yields a color excess comparable to that of the bluest \civ-weak composite, with 0.019$\pm$0.005 (0.047$\pm$0.013).

For further, independent, dust constraints, we also use the Balmer decrement from H$\delta$ and H$\gamma$ emission, where available. We measure H$\delta$/H$\gamma$=0.27$\pm$0.23 and 0.30$\pm$0.17 in the moderately-blue and bluest stacks, respectively, which assuming Case B recombination at $T\approx20,000 K$ are consistent with dust-free ratios of H$\delta$/H$\gamma$=0.27 \citep{osterbrock06}. A similar conclusion can also be derived for the \civ-strong composite, which yields H$\delta$/H$\gamma$=0.50$\pm$0.17. The high value observed in the \civ-strong composite is well above the dust-free limit from Case B recombination theory; similar cases have been observed in low-metallicity extreme emission line galaxies at high redshift using H$\alpha$ and H$\beta$ (e.g., \citealt{endsley21b,pirzkal24,cameron23_nebular,shapley23}), suggesting possible deviations from Case B recombination \citep{scarlata24}. Only [\oii] is detected in the reddest composite, therefore an estimate of the Balmer decrement is not possible. Nonetheless, the measurements presented here suggest the bulk of our $z\geqslant10$ sample, and in particular \civ-strong objects, are consistent with dust-poor or dust-free ISM conditions.

Assuming no dust based on the inferred Balmer decrements, we now assess plausible contributions to the intrinsic scatter of $\beta$ from reasonable assumptions on the nebular continuum, gas-phase metallicities, stellar ages, and star formation history. Using the Flexible Stellar Population Synthesis (FSPS) code \citep{conroy09,conroy10}, we generate a number of simple stellar population (SSP) models and compare their range in $\beta$ to our observed values. The chosen models are: ($i$) Two SSPs fixed at 1\% and 5\% metallicity, including nebular continuum. ($ii$) Two SSPs in a burst of star formation, one in the middle of the burst and one 10 Myr after the burst (with 5\% metallicity and including nebular continuum). ($iii$) Two SSPs at stellar ages of 10 Myr and 100 Myr (with 5\% metallicity and nebular continuum). ($iv$) Two SSPs with 5\% metallicity, one with and one without nebular continuum. We allow each of the aforementioned models except for ($iii$) to evolve with stellar ages from 1 Myr to 300 Myr.

\begin{figure}
\center
\includegraphics[width=\columnwidth]{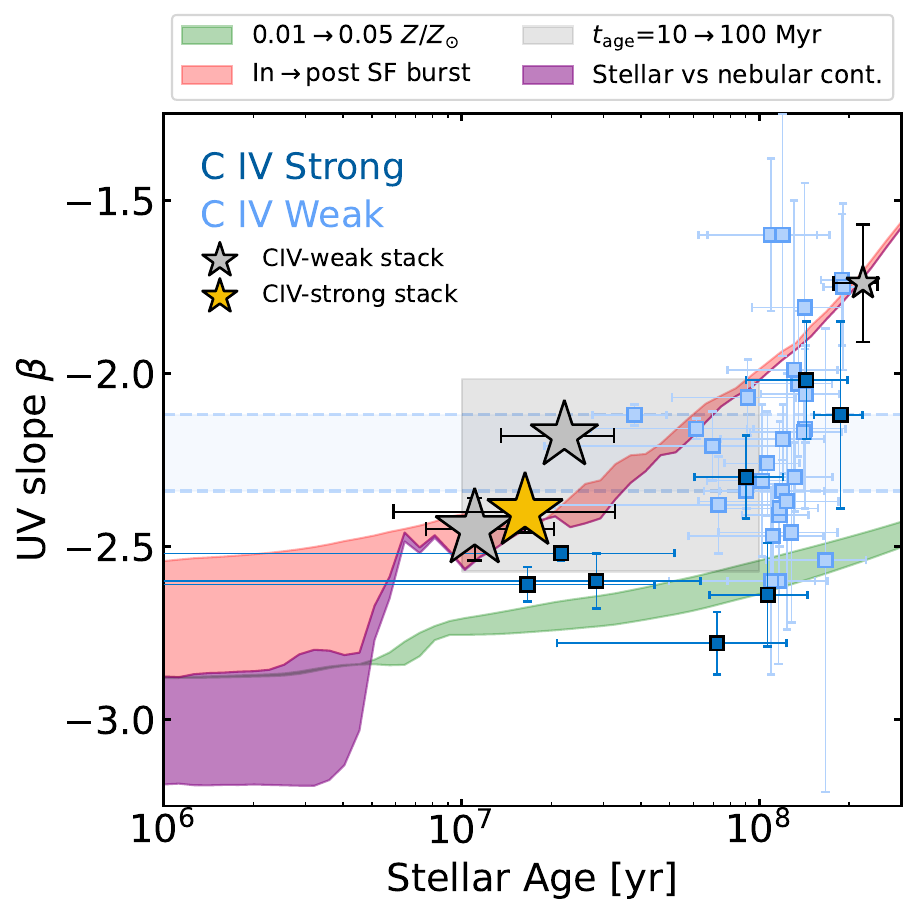}
 \caption{\textbf{Contributions to the UV continuum slopes of $z\geqslant10$ galaxies.} The mass-weighted ages from \texttt{Bagpipes} are plotted with the spectroscopic $\beta$ measurements for individual objects (small squares) and composite measurements (stars), following previous color schemes. Composite symbols are scaled in size according to their recent star formation activity, where a larger size equates to more recent star formation. The intrinsic scatter in $\beta$ measured in Section~\ref{subsec:beta} for \civ-weak objects is shown as the light-blue box, while the range of $\beta$ determined from a number of FSPS models are shown as green fill (varying metallicity), red fill (timing relative to a burst of star formation), purple fill (stellar vs nebular continuum contributions), and grey fill (varying stellar ages). Only variations in the recent star formation history and in stellar age are able to explain the intrinsic scatter of observed UV slopes.}
 \label{fig:beta_origins}
\end{figure}

We plot the distribution of mass-weighted stellar ages from our \texttt{Bagpipes} fits as a function of $\beta$, as well as the ranges of $\beta$ from the FSPS models, in Figure~\ref{fig:beta_origins}. The individual spectra span ages 16-222 Myr, with a median of 72 Myr and 16th-84th percentiles of 116-143 Myr, respectively. Our composite spectra are characterised by ages 222$\pm$18 Myr, 22$\pm$1 Myr, and 11$\pm$5 Myr for the reddest, moderately-blue, and bluest \civ-weak composites, highlighting the association of the bluest slopes with the youngest stellar ages, and vice versa. For comparison, the \civ-strong composite shows a similarly young age of 16$\pm$5 Myr, and a UV slope comparable to the bluest \civ-weak composite. 

Comparing to the FSPS models to explain the range of observed $\beta$ values, we find important differences between models of comparable stellar ages. Metallicity changes from 1-5\% solar are insufficient at all ages to explain the observed (intrinsic) scatter of $\beta$ associated with the \civ-weak sources, while the nebular continuum is only expected to redden the UV slopes of especially young sources ($\lesssim5$ Myr) up to approximately $-2.7$. In contrast, the recent star formation activity can generate up to $\Delta\beta\simeq0.4$ scatter for sources with ages $1-10$ Myr, which is larger than the measured intrinsic scatter. This is comparable to the scatter of $\Delta\beta\simeq0.5$ introduced by varying the stellar age from 10 Myr to 100 Myr, which are comparable to the range of ages seen in our data points. Of course, each of these is intrinsically degenerate: a recent burst of star formation would likely (temporarily) outshine the rest of the galaxy with especially young stars and enhanced nebular continuum, which results in overall blue UV continuum slopes, before returning to its equilibrium state (e.g., \citealt{gelli24}). This is particularly evident in Figure~\ref{fig:beta_origins} through the parameter space occupied by the \civ-strong and bluest \civ-weak composites, as well as through the scaling of their points as a function of recent star formation history: it becomes immediately clear that the highest star formation activity is associated with the youngest stellar ages and bluest slopes, highlighting how newly-formed O and B stars dominate the UV spectrum in recent bursts of star formation. With these comparisons in mind, we can conclude that the bulk of our sample are dominated by young stellar populations with virtually no dust and small metallicity variations, whose observed intrinsic scatter is likely driven by variations in stellar age and recent star formation history.

\subsection{Stochastic Star Formation from Compact Morphologies and Recent Bursts}
\label{subsec:bursty}
\begin{figure}
\center
\includegraphics[width=\columnwidth]{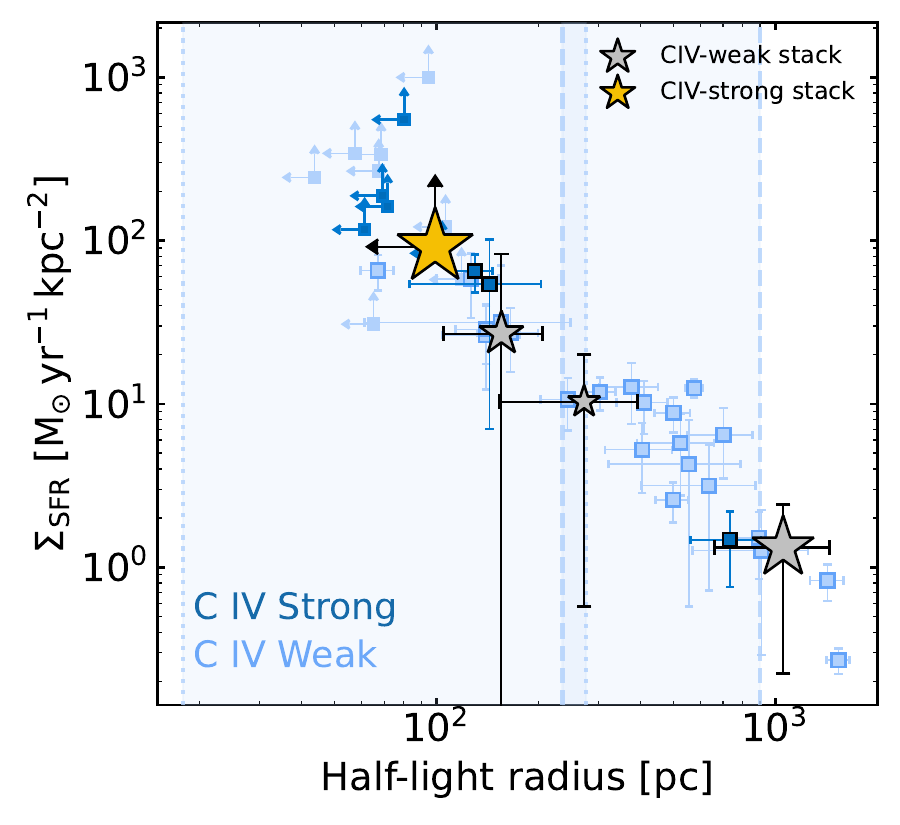}
 \caption{\textbf{The burstiness of star formation in $z\geqslant10$ galaxies.} Individual measurements are plotted as small squares following previous color schemes, while composite measurements from stacked NIRSpec spectra and NIRCam F200W data of \textit{unlensed} sources are shown as stars (grey for the \civ-weak population, gold for the \civ-strong population). Each composite symbol is scaled in size according to their recent star formation activity, where larger size corresponds to more recent star formation. The log-normal scatter around the median of the morphological distributions in Section~\ref{subsec:morphologies} are plotted as light blue boxes. The large range of SFR surface densities and correlation with half-light radius suggests varying gas densities and star formation efficiencies, while the positions of \civ-strong objects at the most compact end of the distribution with high star formation rate surface densities highlights the requirement of extreme star formation episodes to power \civ\ emission.}
 \label{fig:bursty}
\end{figure}

The large range of half-light radii presented in Section~\ref{subsec:morphologies} likely reflects varying gas densities and star formation efficiencies, resulting in an equally large range of star formation activity (e.g., \citealt{morishita24}). Accordingly, the near ubiquity of strong \civ\ emission at the most compact end of the size distribution plausibly points to an association with the most extreme star formation activity. Indeed, \citet{senchyna19} established for a sample of dwarf galaxies with strong \civ\ emission a clear link to very high specific star formation rates in low metallicity environments, while recent work by \citet{schaerer24} found strong \niv] emitters displayed especially high star formation rate and stellar mass surface densities compared to the more general population of sources at $5<z<14$. Here we verify the association between strong UV line emission and extreme star formation activity in our extended $z\geqslant10$ sample, by inferring a star formation rate surface density ($\Sigma_{\rm SFR}$) from the half-light radii presented in Section~\ref{subsec:morphologies} using a SFR estimated directly from the UV continuum and corrected for lensing magnification, following:

\begin{equation}
    \rm{SFR\,[M_{\odot}\,yr^{-1}]} = 1.0\times 10^{-28}\,L_{\rm UV}/\mu\,[erg\,s^{-1}\,Hz^{-1}],
\end{equation}

where the scaling factor is adjusted to $\sim10$\% metallicity and assumes a \citet{kroupa01} IMF. The derived SFR densities are shown as a function of half-light radius in Figure~\ref{fig:bursty}, along with the intrinsic scatter of the latter derived in Section~\ref{subsec:morphologies}.


We observe a large range of $\Sigma_{\rm SFR}$, with resolved measurements for the \civ-weak sample generally spanning $\sim1-93$ $M_{\odot}$\,yr$^{-1}$\,kpc$^{-2}$. These appear, as expected, to be primarily driven by the distribution of half-light radii from Section~\ref{subsec:morphologies}; the most compact objects display the most significant star formation activity, and vice-versa, possibly due to enhanced gas densities and star formation efficiencies. A smaller number of \civ-weak sources (6 objects) also display significantly higher $\Sigma_{\rm SFR}$ values compared to the bulk of the sample, with $>100 M_{\odot}$\,yr$^{-1}$\,kpc$^{-2}$. These are characterised by lower limits governed by unresolved half-light radii, indicating extremely bursty systems. In comparison, the \civ-strong objects show a far smaller distribution of values compared to the full distribution of \civ-weak sources, with all but one lying in regions of particularly high SFR surface densities ($\gtrsim50$ $M_{\odot}$\,yr$^{-1}$\,kpc$^{-2}$) and compact half-light radii, and 4 out of 7 of those sources characterised by lower limits in regions of extreme star formation activity ($\gtrsim100$ $M_{\odot}$\,yr$^{-1}$\,kpc$^{-2}$).

To further anchor these observations, we construct composite spectra and median NIRCam F200W images of \textit{unlensed} \civ-weak objects in bins of half-light radius, namely $r_{\rm 50}<100$ pc (7 sources), $100<r_{\rm 50}<400$ pc (11 sources), and $r_{\rm 50}>400$ pc (13 sources). The same is done with the \civ-strong objects, again excluding lensed sources to avoid artificially biasing size measurements. For the NIRCam stack, each image is normalised by the total flux of the source, as defined by its segmentation map, prior to going into the stack. The resulting sizes are measured according to the procedure in Section~\ref{subsec:morphologies} and we use the median UV luminosity of the composite spectra to derive a SFR surface density. The resulting measurements yield half-light radii of 156$\pm$50 pc, 272$\pm$119 pc, and 1049$\pm$390 pc for the \civ-weak composites and upper limits of $<98$ pc for the \civ-strong stack, together with SFR surface densities of 1$\pm$1 $M_{\odot}$\,yr$^{-1}$\,kpc$^{-2}$, 10$\pm$9 $M_{\odot}$\,yr$^{-1}$\,kpc$^{-2}$, 26$\pm$54 $M_{\odot}$\,yr$^{-1}$\,kpc$^{-2}$ and $>91$ $M_{\odot}$\,yr$^{-1}$\,kpc$^{-2}$, respectively, which are added to Figure~\ref{fig:bursty} and whose symbols are once again scaled in size according to the fraction of total star formation captured in the last 10 Myr using their best-fit \texttt{Bagpipes} SEDs.

The composite measurements clearly reaffirm what is already suggested through the individual measurements, namely that the \civ-weak objects are generally characterised by larger half-light radii and lower star formation activity, while \civ-strong objects reside at the most extreme end of these two distributions with the most compact sizes and most intense star formation activity. This is illustrated both in the parameter space occupied by the composites but also the scaling of their symbol sizes by the recent star formation history, where the \civ-weak composites display reduced star formation activity over the most recent 10 Myr compared to the \civ-strong composite. Morever, the UV slopes associated with the extended, moderately-compact, and compact stacks are $-$2.10$\pm$0.04, $-$2.19$\pm$0.04, and $-$2.21$\pm$0.05, pointing to increasing fractions of progressively younger stellar populations which is expected given the observed trend with SFR surface density. Each of these remain noticeably redder than the UV slope of the \civ-strong composite with $-$2.39$\pm$0.06, corroborating the marked differences in star formation related properties between the two sub-populations. The plot serves as another example of strong \civ\ (and \niv]) emitters' place at the tail-end of morphological and star formation related distributions, corroborating the notion that these sources are powered by intense and bursty star formation episodes. We explore this further in Section~\ref{subsec:sfh}.

\section{What Drives the Diversity of Galaxies Within the First 500 Myr?}
\label{sec:diversity_disc}
The limited scatter in ISM and stellar conditions determined in previous sections, together with overlap between the properties of \civ-weak and \civ-strong objects, points to a common evolutionary pathway regulating the physics of these two sub-populations and our $z\geqslant10$ sample as a whole. The generally smooth nature of the \civ-weak distributions further indicates that the mechanisms governing the spectral and morphological diversity must operate broadly across the population, rather than being confined to distinct subsets, and be capable of reproducing features seen at either extremes and in particular those of \civ-strong sources. This implies either time-dependent processes and/or an additional but widespread characteristic of high-redshift sources, such as early SMBHs or exotic stellar populations. As such, we turn to the assembly histories and AGN nature of our sample, and explore how these factors combine to shape the scatter and emerging picture of galaxy evolution within the first 500 Myr.

\subsection{Short-Lived Extremes Driven by Bursts and Lulls of Star Formation}
\label{subsec:sfh}
Throughout Sections~\ref{sec:distribution} and \ref{sec:props}, our measurements have shown a limited distinction between the observed properties of the \civ-strong objects relative to the distribution of equivalent measures for the \civ-weak sample. This suggests such objects do not represent a fundamentally distinct class from the \civ-weak population, although their location at the tail of most distributions highlights more extreme physical conditions powering their spectra.

Moreover, our measurements have shown a noticeable correlation between the scatter in the $z\geqslant10$ sample and the recent star formation activity deduced from their \texttt{Bagpipes} star formation histories, as seen through the scaling of symbols in Figures~\ref{fig:metal}, \ref{fig:beta_origins}, and \ref{fig:bursty}. This is perhaps not unexpected given recent studies that have inferred burst-driven star formation from observational data, or proposed stochastic activity to explain the excess luminosities seen in $z\geqslant10$ galaxies \citep{mason23,shen23,ciesla24,endsley25,rojasruiz25,tacchella23,gelli24,simmonds25,mcclymont25}. In this regard, a marked difference between the \civ-strong and \civ-weak sub-populations has been the association of the former with strong bursts of recent star formation. Noting the extremely compact morphologies of the \civ-strong population seen in Figure~\ref{fig:morpho}, episodic star bursts would serve as a natural feature of a picture whereby the large range of half-light radii is modulated by outshining effects arising from mergers, accretion events, and/or bound short-lived star clusters.

To better evaluate the possible influence of recent star formation to strong \civ\ emission, we now use \texttt{Bagpipes} to characterise time-dependent origins, through a more detailed look at the star formation histories of the composites derived in the previous section. We define burstiness parameters tracing various timescales, namely $f_{\rm burst,age}=$SFR$_{\rm age}$/SFR$_{\rm tot}$, where SFR$_{\rm age}$ corresponds to the integrated star formation over lookback times of 0-3 Myr, 0-5 Myr, or 0-10 Myr, and SFR$_{\rm tot}$ corresponds to the integrated star formation over the lifetime of the galaxy set by its redshift.

Evaluating first the distribution of $f_{\rm burst}$ for star formation activity within 0-10 Myr, we find 71\% of the sample show SFHs dominated by this recent activity ($>50$\% of the total SFH), in line with expectations given the short lifetimes of the sample ($\simeq16-222$ Myr, see Figure~\ref{fig:beta_origins}). On these timescales, we find no difference in the range of $f_{\rm burst,10 Myr}$ between the \civ-weak and \civ-strong samples, in line with the overlap in parameter space seen Figure~\ref{fig:bursty}. Reducing the burst timescales to 0-5 Myr, differences between the two samples begin to appear. The overall distribution of $f_{\rm burst,5 Myr}$ displays a wider spread and shifts towards lower values, with only 44\% of the sample characterised by $f_{\rm burst,5 Myr}>50$\%. We note that 5/8 of the \civ-strong sources harbour $f_{\rm burst,5 Myr}>60$\%, although a non-negligible fraction of \civ-weak sources ($\simeq20$\%) also show similarly high $f_{\rm burst,5 Myr}$ fractions. The most significant differences appear when considering burst fractions over the most recent timescales of 3 Myrs. In such cases, galaxies with strong \civ\ emission confidently reside at the highest end of the $f_{\rm burst,3 Myr}$ distribution, while \civ-weak sources populate the parameter space further below.

\begin{figure*}
\center
\includegraphics[width=\textwidth]{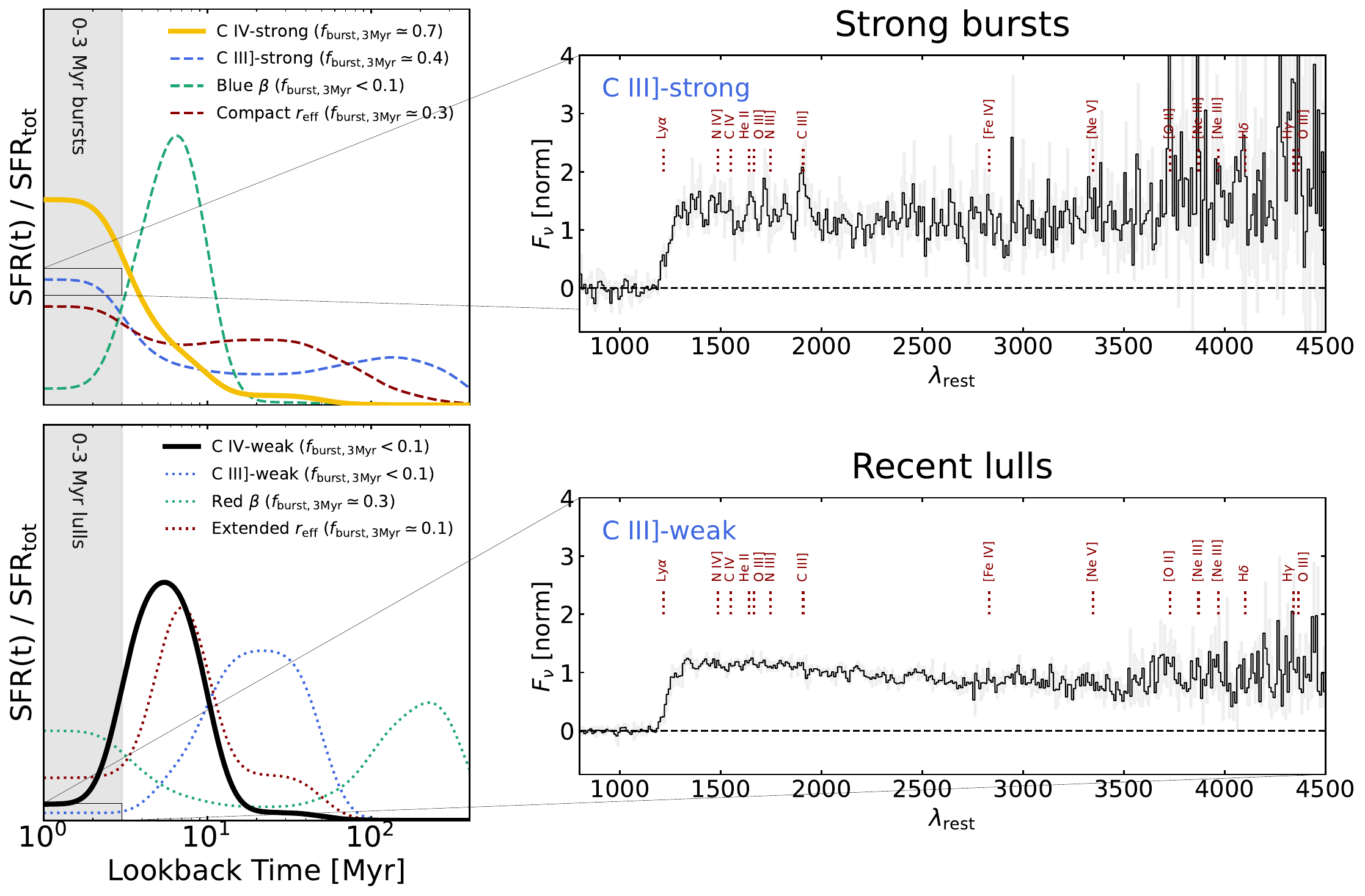}
 \caption{\textbf{The average star formation histories of $z\geqslant10$ galaxies (left) and their resulting spectra (right).} We show these for composite spectra at either end of the distributions discussed in previous sections (properties conducive to the strongest star formation activity in the top left panel, and those conducive to weaker activity in the bottom left panel), as a function of lookback time. The SFHs are smoothed and normalised by the integral of the full SFH, for clarity. Each SFH has a reported burstiness parameter, defined as $f_{\rm burst}$=SFR$_{0-3\,\rm{Myr}}$/SFR$_{\rm tot}$, where SFR$_{0-3\,\rm{Myr}}$ is the star formation history integrated from 0-3 Myr and SFR$_{\rm tot}$ is the integral of the total star formation history across the lifetime of the source set by its redshift. $z\geqslant10$ galaxies showcase a diverse range of SFHs with high stochasticity characterised by recent upturns and significant lulls in star formation within their short lifetimes. This is demonstrated by two composite spectra on the right, chosen to illustrate the effects of a recent burst (the \ciii]-strong composite, top right) and one showing the effects of a recent lull (the \ciii]-weak composite, bottom right). The sharp increase in the most recent star formation history ($\lesssim3$ Myr) also serves as the primary predictor for strong \civ, with such emitters characterised by $f_{\rm burst,3 Myr}\simeq$0.69 in contrast to the more moderate values of 0.01-0.42 for the \civ-weak sources.}
 \label{fig:sfh}
\end{figure*}

This is shown more clearly when comparing across our sub-divided composite spectra. In the left column of Figure~\ref{fig:sfh} we compare the \texttt{Bagpipes}-derived star formation histories for the spectroscopic composites at the extreme ends of distributions linked to star formation (i.e., \ciii]-strong vs \ciii]-weak, bluest $\beta$ vs reddest $\beta$, and compact vs extended half-light radius). As expected we find a variety of star formation histories across the extended lookback times of 0-300 Myr, highlighting the stochastic nature of mass build up in the early Universe. Generally-speaking, composites with properties conducive to star formation show rising SFHs as they evolve and the most enhanced activity on recent timescales of 0-10 Myr, while their less active counterparts show some evidence for star formation episodes at lookback times $>10$ Myr and a downturn in activity at more recent times of 0-10 Myr. Moreover, we find little difference in the burst fraction associated with 0-10 Myr timescales, in line with comparisons across individual measurements. The most marked differences are found on $\leq$3 Myr timescales, where the most active composites generally showcase the most pronounced bursts of star formation within their lifetime, in contrast to their less active counterparts which show clear lulls in star formation activity. This is particularly evident when comparing the two extreme ends of the \ciii] EW distribution, where the \ciii]-strong composite shows traces of recent star formation activity through the presence of UV and optical emission lines, while in contrast the \ciii]-weak composite remarkably shows a complete absence of emission lines across the entire rest-frame UV to near-optical, together with very low escape fractions of ionising photons and a hint of a Balmer break. The spectrum is similar to a number of other sources at $z\simeq4-8$ showcasing significant lulls at the time of observation \citep{looser23,dome24,looser25,endsley25}; given the high redshifts considered, these sources are not expected to maintain such low levels of star formation for long and are likely to undergo additional bursts of star formation (or ``rejuvenation events''; \citealt{witten25}).

The SFH of \civ-strong objects shows by far the most pronounced and significant burst of all the spectra composites within the most recent 3 Myr, with $f_{\rm burst,3 Myr}\simeq0.7$ that is unmatched by the \civ-weak objects (c.f. with $f_{\rm burst,3 Myr}\simeq0.4$ for the \ciii]-strong and smallest $r_{\rm 50}$ composites). The contrast is even more pronounced when comparing to the \civ-weak composite ($f_{\rm burst,3 Myr}\simeq0.1$), which shows most of its star formation activity occurring at lookback times $5-10$ Myr.

Our exercise clearly highlights the bursty nature of early galaxy evolution, and the necessity of modelling star formation on very short timescales to accurately match high redshift spectra. In particular, the comparison shows how strong \civ\ emitters are predominantly a result of very recent and extreme bursts of young star formation, the likes of which are beyond even the most intense \civ-weak populations and represent the primary distinguishing feature between sources with or without strong \civ\ emission. As such, it appears clear that the observed spectral diversity among the $z\geqslant10$ population is driven primarily by the recent star formation history, whereby strong UV line emitters represent brief and extreme snapshots of a common evolutionary pathway.

\subsection{Are Strong \civ-Emitters Also Nitrogen-Enhanced?}
\label{subsec:n-over-o}
A number of recent studies have reported nitrogen-enhancement (relative to oxygen) in objects with strong \civ\ emission (e.g., \citealt{cameron23,castellano24,topping24,naidu25}), hinting at common underlying ISM and stellar properties. Given the extreme conditions characteristic of the strong \civ\ emitters outlined in this study, it is conceivable that strong nitrogen emitters trace similarly extreme conditions and may be subject to the same duty cycle advocated for in the previous section. In such a case, the implications would be that the processes driving such nitrogen enhancement represent a common but brief characteristic among the general $z\geqslant10$ population, visible only in short-lived periods along a source's recent star formation history.

With this in mind, Figure~\ref{fig:stack} already hints at such a scenario through a \civ-\niv] association: the \civ-strong composite shows a clear detection of \niv] emission (as well as a tentative 2.2$\sigma$ \niii] line), while none of the \civ-weak composites show convincing evidence for similar nitrogen emission. To verify the abundances that these line detections and ratios imply, we use \texttt{PyNeb} \citep{luridiana15} to determine ionic abundances of nitrogen and carbon, while using the gas-phase oxygen metallicities derived in Section~\ref{subsec:ciii} for oxygen abundances. We assume an electron density of $n_{\rm e}=30,000$ cm$^{-3}$ from the intermediate-ionisation zone relation of \citet{topping25b}, determined from the \ciii] doublet. Measurements at such high redshifts are scarce (e.g., \citealt{abdurrouf24}), and while electron densities in both low- and high-ionisation have been shown to evolve strongly between $z\simeq0-5$, the evolution appears to plateau at higher redshift and order-of-magnitude deviations from the relation are not expected \citep{li25,topping25b,harikane25b}. A major source of uncertainty comes from a lack of constraints on the electron temperature -- although the resolution at the reddest end of the prism allows for detections of the [\oiii]$\lambda$4363 \AA\ line, strong [\oiii]$\lambda$5008 \AA\ emission is beyond NIRSpec's coverage. We therefore assume a range of values 15,000-20,000 $K$, which is in line with auroral line temperature measurements at $z\simeq9-10$ \citep{sanders_metal,pollock25}, and assume a fiducial value of 17,000 $K$ from the $z\geqslant10$ estimate of \citet{hsiao24b}. Moreover, given the moderate-to-high ionisation state of the detected nitrogen and carbon gas, we consider whether ionisation correction factors (ICFs) are required to capture their full extent and total abundances (note that our derived oxygen abundances already include such corrections and represent total abundances). We therefore compute a range of N/C and C/O ICFs (using the emission lines adopted in our \texttt{PyNeb} analysis) as a function of log\,$U$ and Ne3O2 ratio, based on a grid of CLOUDY and BPASS models with metallicities less than 20\% solar, log\,$U$ up to 0, densities up to 3,000 cm$^{3}$, and stellar ages between 3-10 Myr. The resulting ICFs very clearly asymptote to unity at log\,$U\gtrsim-2$ and Ne3O2$\gtrsim$1, both of which are characteristic of our \civ\-strong stack (see Table~\ref{tab:stacks} and Section~\ref{subsec:metalpoor}) and suggest the fraction of N and C in the singly-ionised state is effectively negligible. These findings are also in broad agreement with the lower-redshift determinations of \citet{berg19a} and \citet{martinez25} for the most extreme ISM conditions. As such, we consider our derived abundances close to total, and do any apply any ICF.

\begin{figure}
\center
\includegraphics[width=\columnwidth]{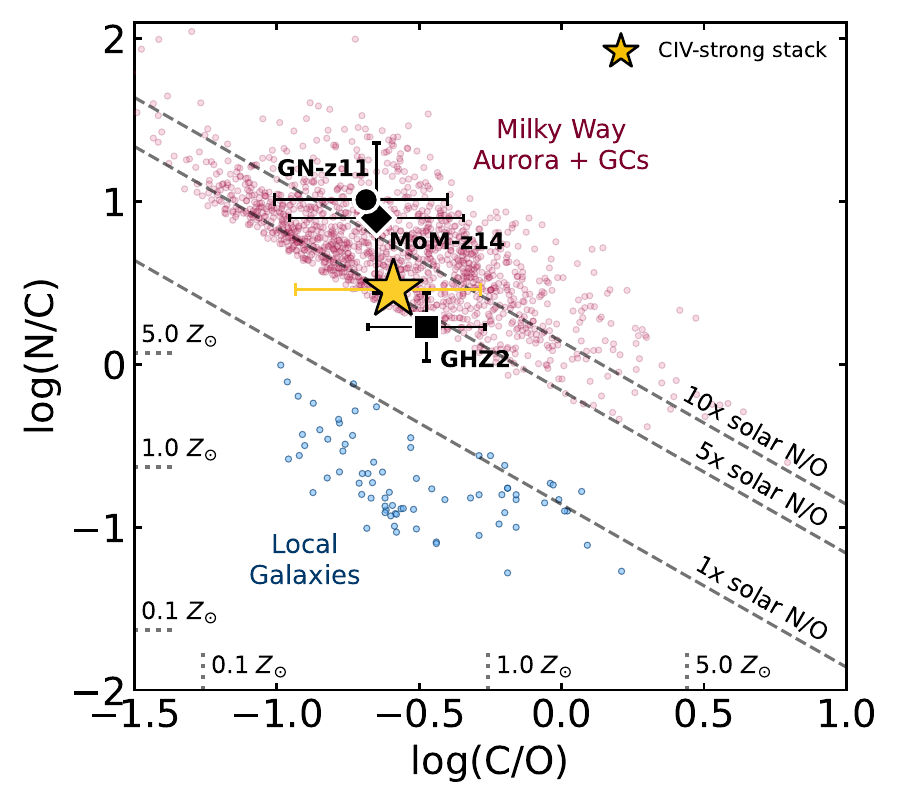}
 \caption{\textbf{Nitrogen-enhancement of strong \civ-emitters.} We show the N/C and C/O abundances for the \civ-strong composite (gold star) and the comparable estimates for $z\geqslant10$ strong \civ-emitters in GN-z11 \citep[][with re-derived O/H estimates from strong line ratios]{bunker23,cameron23_nitrogen}, GHZ2/GLASS-z12 \citep{castellano24}, and MoM-z14 \citep{naidu25} as black symbols. Lines of constant N/C, C/O, N/O abundances are plotted as dotted and dashed lines. The highly super-solar N/C and N/O values point to significant nitrogen enhancement, at odds with local samples (blue points from \citealt{izotov23}) and comparable to nitrogen-enriched globular cluster stars around the Milky Way (``Aurora stars'' in red from \citealt{belokurov23,belokurov24}). The N-enhancement in \civ-strong sources suggests a possible generic mode of early star formation, regulated by the same duty cycle implied by Figure~\ref{fig:sfh} and possibly underpinned by very massive or supermassive stars.}
 \label{fig:logNO}
\end{figure}

With the above considerations in mind, we plot the results of the inference for the \civ-strong composite in Figure~\ref{fig:logNO}, where we show N/C versus C/O abundances and lines of constant N/O abundances relative to solar (assuming solar abundances of log(N/O)$_{\odot}$=$-$0.86, log(C/O)$_{\odot}$=$-$0.26, and log(N/C)$_{\odot}$=$-$0.60; \citealt{asplund09}). Given strong [\oiii]$\lambda\lambda$4960,5008 lines are shifted outside of the NIRSpec spectrum, a major uncertainty in determining relative nitrogen enrichment is the derived oxygen abundance to which it is compared. As shown in \citet{naidu25}, we can bypass this uncertainty by comparing nitrogen abundances relative to carbon, both of which are constrained from detected emission lines. For C/O inferences we utilise the best-fit \texttt{Bagpipes} oxygen abundances from Section~\ref{subsec:ciii} together with the inferred \texttt{PyNeb} carbon abundances from \ciii] and \civ. For comparison, we also plot the relative abundances for the sample of local galaxies from \citet{izotov23}. We find the \civ-strong composite to be highly nitrogen-enriched, with super-solar values of log(N/C)$\simeq0.4-0.5$ ($\sim$10.0-12.6$\times$ solar) at approximately sub-solar C/O abundances (log(C/O)$\simeq-0.9$ to $-0.3$, or $\sim0.2-0.9$ solar), and implied log(N/O) abundances of $\simeq-0.5$ to $0.2$ ($\sim2.3-11.5$ solar). The abundances corresponding to the fiducial electron temperature assumptions are log(N/C)$\simeq0.46$ (11.5$\times$ solar), log(C/O)$\simeq-0.59$ (0.46$\times$ solar) and log(N/O)$\simeq-$0.13 (5.4$\times$ solar).

The derived values are consistent with independent measurements in the literature for the nitrogen abundances in GN-z11, GHZ/GLASS-z12, and MoM-z14 \citep{cameron23_nitrogen,castellano24,naidu25}, all of which contribute towards our composite spectrum. Each of these are characterised by reported log(N/O) abundances of $>-0.25$ ($>4\times$ solar), $-0.29$ to $-0.2$ (4$\times$ solar), and $0.29$ (14$\times$ solar), respectively. Given the more stringent MIRI constraints on the oxygen abundance of GN-z11 from \citet{alvarez_miri}, we re-derive the abundances for GN-z11 using the aforementioned constraints and the line fluxes from \citet{bunker23}, yielding log(N/O)$\simeq$0.32. We plot these and the reported values for GHZ2/GLASS-z12 and MoM-z14 in Figure~\ref{fig:logNO}, for comparison. Although determining a direct association between \civ\ and \niv\ emission will require deeper and higher resolution spectroscopy, it seems clear that both of these high-ionisation lines require similarly extreme conditions for their excitation.

Moreover, given the case for strong \civ\ being the consequence of the extreme peak of a duty cycle, the results shown here suggest nitrogen-enrichment may also be a consequence of the same duty cycle and thus reflective of a generic mode of star formation in metal-poor galaxies at the earliest times \citep{senchyna19,mcclymont25c}. Notably, globular clusters (GCs) have been suggested as one viable mechanism for such enrichment (e.g., \citealt{charbonnel23,senchyna24,rui24}). For instance, \citet{naidu25} compared the nitrogen enhancement of MoM-z14 to the super-solar abundances found in some present-day GC stars and field stars touted as relics of early in situ Milky Way star formation. In this scenario, these nitrogen-rich and metal-poor stars are thought to have formed prior to the disk component of the Milky Way (i.e., in its ``Aurora'' component; \citealt{conroy22,belokurov22,semenov24}) within dense and massive star clusters similar to GCs, and thus may have contributed significantly to the past star formation of the Milky Way at higher redshift. A sample of these GC and ``Aurora'' stars from \citet{belokurov23,belokurov24}, selected a priori to have high N/O abundances, are plotted in Figure~\ref{fig:logNO}, together with the local galaxy sample from \citet{izotov23}, for comparison.  A physical picture where GCs and/or their precursors represent the source of nitrogen enrichment at high redshift brings together a number of key considerations, namely: (i) A natural duty cycle where the possible presence of very massive stars (VMS) or supermassive stars (SMS) temporarily enhances nitrogen abundances relative to oxygen, before leaving behind a wake of more standard abundance patterns from H-burning in the CNO cycle. (ii) Reduced mass-to-light ratios from a top-heavy IMF which significantly reduces the tension between theoretical predictions and observational results at the bright end of the $z\geqslant10$ UV luminosity function. (iii) A natural morphological dichotomy such as that observed in Section~\ref{subsec:morphologies}, whereby the light and mass in more compact objects is potentially dominated by only a few massive GCs that outshine the rest of the host galaxy. A number of confirmed high-redshift sources benefitting from gravitational lensing have already shown examples, and even the prevalence, of resolved star clusters at $z\sim6-10$ \citep{vanzella23,vanzella25,fujimoto24,adamo24,claeyssens25}, highlighting the feasibility of this interpretation.

\subsection{Could SMBHs Play a Role?}
\label{subsec:agn}
\begin{figure}
\center
\includegraphics[width=\columnwidth]{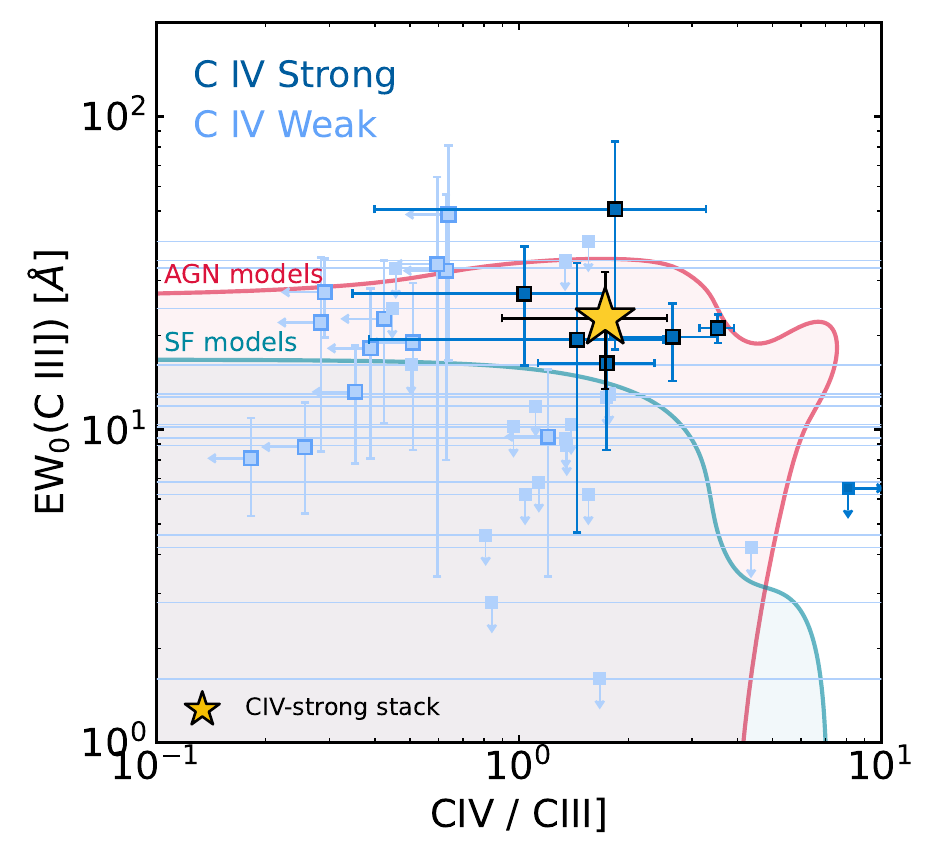}
 \caption{\textbf{Star formation vs SMBH activity in $z\geqslant10$ galaxies.} We plot measurements of \ciii] EW and \civ/\ciii] ratios for our sample of sources (individual measurements as squares, following previous color schemes) and the \civ-strong composite (gold star), and compare these to the photoionisation models for star-forming galaxies (green) and AGN (red) of \citet{nakajima22} (shown as contours encapsulting 95\% of the data points). The uncertainties of our data points and the significant overlap between models prevent us from ruling out either one of the photoionisation origins, however at least $\sim$50\% of our sample are most consistent with star formation based on their \ciii] EW upper limits.}
 \label{fig:agn}
\end{figure}

Thus far we have argued for time-dependent star formation processes as the primary explanation for the strong \civ\ and \niv] emission. However, given the revelation of significant numbers of low-luminosity AGN at lower redshifts (e.g., \citealt{maiolino23,greene23,scholtz25a}), we explore whether the $z\geqslant10$ spectra studied here display any compelling evidence for significant AGN activity. We consider EW$_{0}$(\ciii]) and \civ/\ciii] as metrics to separate star formation from AGN activity \citep{nakajima22,castellano24,napolitano25}, given the high cadence of \ciii] emission in our sample and the absence of rest-frame optical diagnostics. Using line measurements from Section~\ref{subsec:measurements}, we compare to star formation and AGN models from \citet{nakajima22} with similar ionisation parameters and gas-phase metallicities to those reported in Section~\ref{subsec:ciii} (we note the degree of overlap between models is highly sensitive to these assumptions). The results are shown in Figure~\ref{fig:agn}, where sources with \ciii] detections are found to populate areas of the parameter space consistent with either population (within uncertainties), as has become a common theme at high redshift. Only a small range of \ciii] EW clearly distinguishes between the two models (with $\gtrsim16.6$ \AA\ predominantly occupied by AGN models), and we find 14 \ciii]-detected sources -- including all but one strong \civ\ emitter -- lie above the demarcation in the parameter space characteristic of AGN, although all but 5 of these remain consistent (within uncertainties) with photoionisation by star formation. All other sources in our sample are characterised by \ciii] upper limits that fall below the star-forming demarcation and are thus inconsistent with AGN activity based on this metric.

Even for the highest EW$_{0}$(\ciii])s reported here, however, we note that unambiguous AGN claims with these tracers are challenging given the relatively low excitation energy of \ciii] and the comparable EWs reported for star-forming sources at $z\sim5-10$ (c.f. with \citealt{rb24,tacchella25}). Indeed the AGN nature of a number of the sources in our sample has already been independently studied with a more extensive suite of photoionisation models, each yielding comparable conclusions (e.g., \citealt{castellano24,napolitano25,napolitano25b}). The comparison is emblematic of the difficulty in isolating AGN activity from the high ionisation conditions that result from the low metallicities and hard radiation fields at high redshift; for instance, even \heii$\lambda$1640 \AA, with an excitation energy of 54.4 eV, is found to be prevalent in stacked spectra of $z\sim7$ star-forming systems using prism and $R\sim1000$ data \citep{hu24,glazer25}. Detections of extremely high ionisation lines (with $>>50$ eV) such as \nv\ emission (77.5 eV) provide one alternative signature which is extremely challenging to reconcile with star formation alone and provide a more direct tracer of AGN activity \citep{feltre16,laporte17b,mainali18}, however we find no clear evidence of such a line in any of our spectra (individual or composites). The identification of broad components in Balmer lines, paired with their absence in forbidden lines, suggests the presence of a broad-line region (BLR) and provides another conclusive alternative (e.g., \citealt{matthee24,morishita25,alvarez_miri}). However, the faintness of H$\delta$ and $H\gamma$ in our spectra make such identification impractical.

In Figure~\ref{fig:beta} we found a small number of sources displaying especially red UV continuum slopes, the likes of which are unlikely to be a result of strong nebular continuum and instead require either significant dust obscuration which contrasts with the bulk of the population, or AGN contributions. Indeed two $z\sim10$ sources in our sample, GHZ9 and UNCOVER-26185 (or ``UHZ1''), have already been reported to showcase both star formation and AGN activity based on the detection of a number of high-ionisation UV lines and a potential X-ray detection \citep{kovacs24,napolitano25b,goulding23,bogdan24}, although these and similar claims are contested based on rest-optical measurements from MIRI spectroscopy and imaging \citep{alvarez_miri,alvarez_miri2,zou26}. No Balmer lines or high ionisation UV lines except for \ciii] (EW$_{0}$(\ciii])$\sim13.1\pm6.5$ \AA) are detected in the composite spectrum of those red sources, however, and thus further data are required to discern their nature.

As such, while we cannot rule out minor AGN contributions to the spectra of individual \ciii]-detected sources (see e.g., \citealt{arevalogonzalez25}), the vast majority of these remain consistent with photoionisation by star formation. A similar conclusion can be drawn for the other half of the sample without \ciii] detections, whose \ciii] EW upper limits are unable to reach the high values more characteristic of AGN activity. Since the scatter in \ciii] EWs reported in Section~\ref{fig:ciii} is consistent with variations in recent star formation history, we deem an AGN interpretation as the primary driver for high EWs as only a potential secondary explanation given the lack of standout evidence for such contributions.

\section{Summary \& Conclusions}
\label{sec:summary}
The remarkable diversity of spectroscopic signatures seen within the first 500 Myr of the Universe, hinted at by early NIRSpec observations of $z\geqslant10$ galaxies and now confirmed through the enlarged sample and analysis presented here, suggests a variety of astrophysical conditions. Discerning whether those conditions (as traced by the prominence or absence of high ionisation UV lines \ciii], \civ, and \niv]) trace distinct and separate galaxy populations, or temporary snapshots along a common evolutionary pathway, represents a key question to understand the earliest stages of galaxy formation and evolution. Here we have characterised the distribution and scatter of $z\geqslant10$ galaxy properties using individual and composite spectra from the largest compilation of confirmed sources yet, comparing the bulk of the population (the ``\civ-weak'' population) to sources with strong \civ\ line emission (the ``\civ-strong'' population). Our main findings are the following:

\begin{itemize}
    \item The \civ-weak population exhibits a broad distribution of \ciii] EWs spanning $\sim$1-51 \AA, with an intrinsic scatter well-described by a log-normal distribution ($\mu=8.3^{+2.9}_{-2.6}$ \AA\ and $\sigma_{\rm ln}=2.5^{+1.3}_{-0.7}$ \AA). Composite spectra and empirical calibrations suggest these lines trace metal-poor gas with metallicities $\sim1-8$\% solar, indicative of rapid chemical enrichment within only a few ten to hundred Myr. Comparing to tracers of star formation activity, the variation in \ciii] strength is best explained by differences in recent star formation rather than modest metallicity differences, with strong evidence for recent bursts and relative inactivity.
    \\
    \item The same \civ-weak population exhibits a range of UV continuum slopes characterised by a log-normal distribution with $\mu=2.23^{+0.03}_{-0.03}$ and $\sigma=0.11^{+0.04}_{-0.03}$. The median and scatter is consistent with predominantly dust-poor (or dust-free) and young stellar populations including contributions from the nebular continuum. A small fraction of sources lie outside this distribution with either especially red or blue slopes, indicating possible AGN contributions and higher escape fractions of ionising UV photons, respectively. Balmer decrements measured from composite spectra confirm the dust-poor nature of the bulk population.
    \\
    \item The sample reveals a bimodal size distribution with both extended ($r_{\rm 50}\sim200-1000$ pc) and ultra-compact ($r_{\rm 50}\lesssim100$ pc) morphologies. Each component of the bimodal distribution is well-described by overlapping normal distributions parametrised by $\mu_{\rm extended}=460^{+127}_{-325}$ pc and $\mu_{\rm compact}=70^{+278}_{-44}$ pc, with $\sigma_{\rm ln,extended}=0.7^{+0.7}_{-0.3}$ pc and $\sigma_{\rm ln,compact}=1.4^{+0.9}_{-0.7}$ pc. The morphologies are consistent with the interpretation of stellar disk-dominated and merger-dominated systems, respectively, within which outshining effects regulate the scatter.
    \\
    \item Galaxies with the most compact morphologies exhibit the highest star formation rate surface densities ($\Sigma_{\rm SFR}\simeq30-400$ $M_{\odot}$\,yr$^{-1}$\,kpc$^{-2}$) and rising star formation histories where a recent (0-10 Myr) burst dominates the SFH. Extended sources are characterised by lower star formation rate surface densities underpinned by lulls in the recent star formation history, highlighting the stochastic nature of star formation implied by the low $\Sigma_{\rm SFR}$ values and associated \ciii] EW upper limits.
    \\
    \item The \civ-strong population sits at the extreme tail of several of the distributions, characterised by the bluest UV slopes ($\beta\lesssim-2.5$), the most compact morphologies ($r_{\rm 50} \lesssim 100$\,pc), and the highest star formation rate surface densities ($\Sigma_{\rm SFR} \gtrsim 100\,M_\odot\,\mathrm{yr}^{-1}\,\mathrm{kpc}^{-2}$). However, the overlap with a number of \civ-weak sources suggests they are not fundamentally different systems but rather a short-lived phase within a continuous life-cycle of activity. SED-modelling of their star formation histories confirms their properties can be explained with intense bursts of star formation on very short timescales ($<3$ Myr) which contrasts with the weaker activity and recent downturns seen for their \civ-weak counterparts. This naturally  supports their interpretation as sources seen during a transient phase within a shared evolutionary pathway.    
    \\
    \item Strong \niv] emission is also detected in a composite of strong \civ\ emitters, yielding super-solar N/C and N/O ratios at sub-solar C/O abundances. The line is not detected in any other composite, highlighting shared astrophysical conditions and a possible association between strong \civ\ and \niv\ emitters. By extension, this suggests both features are modulated by the same burst-driven duty cycle, which we speculate is concurrent with the presence of nitrogen-enriched massive and/or supermassive stars in globular clusters.
    \\
    \item The majority of the sample is consistent with star formation-driven ionisation based on a comparison of their \ciii] EWs and \civ/\ciii] ratios to photoionisation models, with only a small number of \ciii]-detected sources displaying sufficiently high values which could be AGN-driven. A small subset of five objects show especially red UV continuum slopes which would require either significant dust obscuration in contrast to the bulk of the sample, or AGN activity. As such, the early onset of some SMBH activity in these sources cannot be fully ruled out.
\end{itemize}

The unprecedented spectroscopic capabilities of JWST have now made the exploration and study of the earliest sources practical. The first examples of these hinted at a curious diversity of spectroscopic properties, the likes of which provide strong constraints on the evolutionary pathways available to the first sources. Here we have taken a significant step forward by leveraging the largest sample of confirmed $z\geqslant10$ sources yet to explore this diversity, placing some of the first constraints on the distribution and intrinsic scatter of ``typical'' galaxies at the bright end of the UV luminosity function with the aim of placing the most remarkable sources into context. The improved spectroscopic constraints point to a scenario in which galaxies oscillate through phases of especially low and high star formation activity, and which serve as the primary driver for the spread in observed galaxy properties. Strong UV line emitters in particular appear to represent the apex of such extreme and recent bursts, marking temporary peaks that result in unusually high \civ\ strengths with possible nitrogen-enhancement before fading back towards more moderate levels of star formation. Taken together, the results suggest that galaxies within the first 500 Myr were not governed by smooth and monotonic paths, but rather by rapid fluctuations in star formation, gas accretion, and feedback, creating a heterogenous but interconnected population.

Future studies will require both higher spectral resolution (e.g., NIRSpec $R\sim1000$ and $R\sim2700$ gratings) and longer wavelength (e.g., MIRI) spectroscopy, with which to deliver key gas-phase, stellar, and AGN diagnostics (e.g., the \ciii]$\lambda\lambda$1907,1909 \AA\ and [\oii]$\lambda\lambda$3727,3729 doublets for electron densities, auroral [\oiii]$\lambda\lambda,1661,1666$ \AA\ and/or [\oiii]$\lambda4363$ \AA\ and [\oiii]$\lambda5008$ for direct O/H metallicities, multiple Balmer lines for accurate Balmer decrements and SFRs, Balmer break and rest-optical continuum for accurate stellar ages and masses, line ratios and broad-line features for AGN identification) that are generally beyond prism spectroscopy of the rest-frame UV. While such examples remain exclusive to a limited number of luminous sources, observations over larger $z\geqslant10$ samples are beginning to emerge and will provide invaluable constraints with which to unlock the full distribution and scatter of galaxy properties in the aftermath of the Big Bang.


\section*{Acknowledgements}
We thank the anonymous referee for useful and constructive feedback which improved the manuscript.
GRB is grateful to Vasily Belokurov and Sarah Kane for providing the relevant abundances for the Aurora data in Figure~\ref{fig:logNO}, as well as to Tiger Yu-Yang Hsiao for helpful discussions regarding the MACS 0647-JD source. We are also grateful to Gabe Brammer for useful discussions and his continuous efforts in maintaining and improving the \texttt{msaexp} code, from which the high-$z$ community continues to benefit greatly. Lastly, we also thank the numerous teams of the observational programs used in this study, for developing these valuable data sets. The data used in this study derive from the following programs: 1181 (PI Eisenstein; \citealt{eisenstein_jades}), 1210 (PI Luetzgendorf; \citealt{eisenstein_jades}), 1211 (PI Isaak; \citealt{maseda24}), 1286 (PI Luetzgendorf; \citealt{eisenstein_jades}), 1287 (PI Isaak; \citealt{eisenstein_jades}), 1345 (PI Finkelstein; \citealt{finkelstein25}), 1433 (PI Coe; \citealt{hsiao24}), 2561 (PI Labb\'e; \citealt{bezanson22}), 2750 (PI Arrabal Haro; \citealt{arrabal23}), 3073 (PI Castellano; \citealt{castellano24}), 3215 (PIs Eisenstein \& Maiolino; \citealt{eisenstein23}), 5224 (PIs Oesch \& Naidu; Oesch et al. in prep), 6368 (PI Dickinson; \citealt{kokorev25}). The authors acknowledge the aforementioned teams and PIs where development of their observing program(s) was done with a zero-exclusive-access period.

This work is based on observations made with the NASA/ESA/CSA James Webb Space Telescope. The data were obtained from the Mikulski Archive for Space Telescopes at the Space Telescope Science Institute, which is operated by the Association of Universities for Research in Astronomy, Inc., under NASA contract NAS 5-03127 for JWST. The specific observations analyzed can be accessed via DOI \href{https://doi.org/10.17909/jqj3-ws37}{10.17909/jqj3-ws37}. Some of the data products presented herein were retrieved from the Dawn JWST Archive (DJA). DJA is an initiative of the Cosmic Dawn Center (DAWN), which is funded by the Danish National Research Foundation under grant DNRF140.

RSE acknowledges generous financial support from the Peter and Patricia Gruber Foundation. YF acknowledges supports from JSPS KAKENHI Grant Numbers JP22K21349 and JP23K13149. This work has received funding from the Swiss State Secretariat for Education, Research and Innovation (SERI) under contract number MB22.00072, as well as from the Swiss National Science Foundation (SNSF) through project grant 200020\_207349.

\section*{Data Availability}
The data presented in this paper are derived from JWST NIRSpec/MSA observations that are publicly available in the Mikulski Archive for Space Telescopes (MAST). The reduced spectra are available upon reasonable request to the authors.



\bibliographystyle{mnras}
\bibliography{export-bibtex} 
\clearpage




\appendix
\renewcommand{\thefigure}{A\arabic{figure}}
\setcounter{figure}{0}
\section{Redshift Confirmations of $\lowercase{z}\geqslant10$ Sources}
\label{sec:photoz}
\begin{figure}[H]
\center
\includegraphics[width=0.45\textwidth]{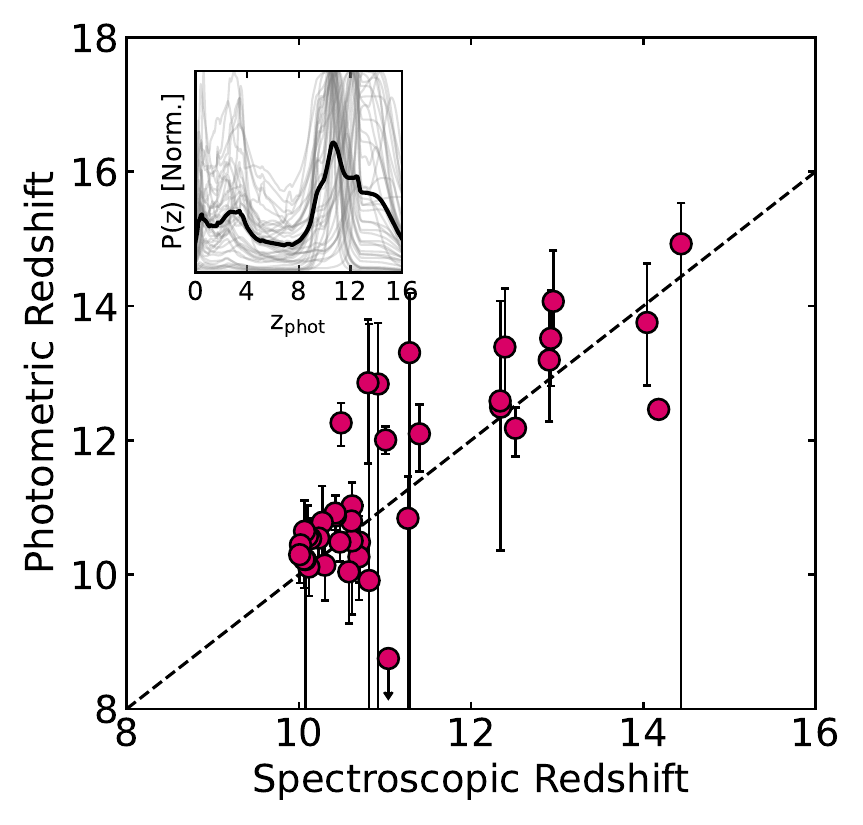}
 \caption{\textbf{A comparison between photometric and spectroscopic redshifts for our sample of NIRSpec-confirmed $z\geqslant10$ objects.} The photometric redshifts were derived using \texttt{EAzY} and the photometric templates of \citet{larson22_eazy}. The photometric redshift solutions for our $z\geqslant10$ clearly favour high-$z$ solutions, adding confidence to those spectroscopic redshifts based on Lyman-$\alpha$ break features. A one-to-one relation is shown as a dashed black line, while the full $P(z)$ distribution for each source is shown in grey in the inset axis and an average $P(z)$ in black.}
 \label{fig:photoz}
\end{figure}

In the absence of clear emission lines, the generally modest depths of NIRSpec prism spectra leave the interpretation of a Lyman-$\alpha$ break open to confusion with a strong Balmer break from low-redshift interlopers. As such, for consistency it is informative to compare the derived spectroscopic redshifts to the photometric redshifts inferred from deeper photometric data. As a sanity check we therefore fit for each source in our sample the HST/WFC3+ACS and NIRCam wide-band photometry from the catalogs of \citet{weibel25}, using the \texttt{EAzY} \citep{brammer08} photometric redshift code. We adopt the theoretical high-$z$ templates of \citet{larson22_eazy} without \lya\ emission (specifically \texttt{tweak\_fsps\_QSF\_12\_v3\_newtemplates}), allowing all combinations between the templates. We assume attenuation of the UV spectrum by the intergalactic medium according to the prescription by \citet{inoue14}, and allow a redshift range $z_{\rm phot}=0-20$ in steps of $\Delta z=0.1$. The resulting $P(z)$ and photometric redshifts are compared to the spectroscopic redshifts in Figure~\ref{fig:photoz}. In all but one case, the preferred photometric redshift solution lies at $z\geqslant10$ and is in very good agreement with the derived spectroscopic redshift, adding confidence to the high-$z$ interpretation despite non-negligible $P(z)$ peaks at lower redshifts of $z\sim2-4$ (see \citealt{castellano25} and \citealt{gandolfi25} for discussions on the nature of low-redshift interlopers). For the one source with a preferred low-redshift solution, CEERS-10, the high redshift nature is unambiguous due to a clear detection of the [\oii]$\lambda\lambda$3727,3729 \AA\ doublet as reported in \citet{arrabal23}. The comparison and $z\geqslant10$ solution for 41/43 spectra in Table~\ref{tab:confirmations} adds confidence in both the photometric pre-selection of high redshift sources as well as the $z\geqslant10$ interpretation of their spectra.

A majority of our sample have furthermore had spectroscopic redshifts reported in previous studies. Comparing our $z\geqslant10$ redshift measurements to those presented in the references in Table~\ref{tab:confirmations}, we report a median difference of $\Delta z=0.02$ and standard deviation of $0.12$, in excellent agreement with previous measurements. Despite the overall good agreement, seven such measurements differ by $\Delta z>0.2$ from ours. Two of these relate to GS-z13-0, who had redshifts of $z=13.2$ and $z=13.22$ reported by \citet{hainline24} and \citet{tang25}, respectively. The absence of measurable emission lines makes the redshift uncertain and largely based on the location of the Lyman-$\alpha$ break; we find our marginally lower redshift of $z=10.926$ to be in slightly better agreement with the location of the break based on our reductions, however deeper NIRSpec or longer-wavelength MIRI spectroscopy will be required to refine the redshift further. Another two relate to UNCOVER-37126, which was measured at $z=10.255$ and $z=10.39$ by \citet{fujimoto24a} and \citet{tang25}, respectively, again based on a fit to the Lyman-$\alpha$ break. Here we detect clear \ciii] emission which refines the redshift to $z_{\rm \ciii]}=10.019$. The remaining measurements with differences of $\Delta z>0.2$ come from a comparison to \citet{tang25} for sources CEERS-10 \citep{arrabal23}, GS-z10-0 \citep{curtislake23}, and CEERS-80041. CEERS-10 was confirmed through the detection of [\oii]$\lambda\lambda$3727,3729 \AA\ and a Lyman-$\alpha$ break at $z_{\rm [\oii]]}=11.043$ by \citet{arrabal23}, and our redshift of $z=11.04$ (c.f. with $z=11.39$ from \citealt{tang25}) is in agreement with this measurement. Similarly, GS-z10-0 was confirmed at $z>10$ through the Lyman-$\alpha$ break by \citet{curtislake23}, but has not yet been verified through clear line emission. Our $z=10.075$ measurement is slightly lower than those reported by \citet{curtislake23} and \citet{tang25} ($z=10.37$ and $z=10.39$, respectively), owing to very marginal but possible detections of [\oii] and [\neiii]$\lambda$3968 \AA. Such tentative lines will have to be confirmed with deeper NIRSpec or far-infrared ALMA spectroscopy. CEERS-80041 is also confirmed through a Lyman-$\alpha$ break due to a relatively shallow and featureless spectrum. There are a number of flux spikes at the red end of the spectrum which could plausibly represent emission lines in the rest-frame optical, however these are far too marginal to be used in a redshift determination. As such, the difference of $\Delta z=0.22$ between our redshift measurement ($z=10.01$) and that of \citet{tang25} ($z=10.23$) is attributed to uncertainties around the Lyman-break. Lastly, we note that one source from the sample of \citet{tang25}, JADES-GS-20021387, is not included in our sample due to flux blueward of the putative Lyman-$\alpha$ break which reveals a low-redshift interloper at a redshift of $z=2.84$.


\bsp	
\label{lastpage}
\end{document}